\documentclass[prd,superscriptaddress,eqsecnum,
nofootinbib,notitlepage,showpacs,preprintnumbers]{revtex4-1}

\usepackage{mathrsfs}
\usepackage{units}
\usepackage{hyperref,graphicx,amsfonts,amssymb,amsthm,amsmath,psfrag}
\usepackage[toc,page]{appendix}
\usepackage{subfigure}
\usepackage{float}
\newcommand{\be}[1]{\begin{equation} \centering \label{#1}}
\newcommand{\ee}{\end{equation}}

\begin{document}
\title{Higgs-Dilaton Cosmology: an effective field theory approach}
\author{Fedor Bezrukov}
\email{fedor.bezrukov@uconn.edu}
\affiliation{Physics Department, University of Connecticut, Storrs, 
CT 06269-3046, USA}
\affiliation{RIKEN-BNL Research Center, Brookhaven 
National Laboratory, Upton, NY 11973, USA}
\author{Georgios K. Karananas}
\email{georgios.karananas@epfl.ch}
\affiliation{Department of Physics,
National Technical University of Athens,
Zografou Campus, 15780 Athens, Greece}
\affiliation{Institut de Th\'{e}orie des Ph\'{e}nom\`{e}nes Physiques, 
\'{E}cole Polytechnique F\'{e}d\'{e}rale de Lausanne, CH-1015 Lausanne, 
Switzerland}
\author{Javier Rubio}
\email{javier.rubio@epfl.ch}
\affiliation{Institut de Th\'{e}orie des Ph\'{e}nom\`{e}nes Physiques, 
\'{E}cole Polytechnique F\'{e}d\'{e}rale de Lausanne, CH-1015 Lausanne, 
Switzerland}
\author{Mikhail Shaposhnikov}
\email{mikhail.shaposhnikov@epfl.ch}
\affiliation{Institut de Th\'{e}orie des Ph\'{e}nom\`{e}nes Physiques, 
\'{E}cole Polytechnique F\'{e}d\'{e}rale de Lausanne, CH-1015 Lausanne, 
Switzerland}
\date{\today}

\preprint{RBRC 1007}

\begin{abstract}
The Higgs-Dilaton cosmological model is able to describe
simultaneously an inflationary expansion in the early Universe and a
dark energy  dominated stage responsible for the present day
acceleration. It  also leads to a non-trivial relation between  the
spectral tilt of scalar  perturbations $n_s$  and the dark energy
equation of state $\omega$.  We study the self-consistency of this
model from an effective field theory point of view. Taking into
account the influence of the dynamical background fields, we determine
the effective cut-off of the theory,  which turns out to be
parametrically larger than all the relevant energy scales from
inflation to the present epoch. We finally formulate the set of
assumptions needed to estimate the amplitude of the quantum
corrections in a systematic way and show that the connection between
$n_s$ and $\omega$  remains unaltered if these assumptions are
satisfied.
\end{abstract}

\maketitle

\section{Introduction}
The shortcomings of the hot big bang model can be solved in an elegant
way if we assume that the Universe underwent an inflationary period in
its early stages. The easiest way for this paradigm to be realized is
by a scalar field slowly rolling down towards the minimum of its
potential \cite{history}. 

 As discussed in  Ref.~\cite{Bezrukov:2007ep}, inflation does not
necessarily  require the existence of a new degree of freedom. The
role of the inflaton  can be played by the Standard Model (SM) Higgs
field with its mass lying in the interval where the SM can be
considered a consistent effective field theory up to the inflationary
scale. More precisely, if the Higgs boson is non-minimally coupled to
gravity and the value of the corresponding coupling constant $\xi_h$
is sufficiently large, the model is able to provide a successful
inflationary period followed by a graceful exit to the standard hot
Big Bang theory \cite{Bezrukov:2008ut,GarciaBellido:2008ab}. The
implications of this scenario have been  extensively studied in the
literature \cite{Bezrukov:2008ej,DeSimone:2008ei,
Barvinsky:2009fy,Barvinsky:2008ia,Bezrukov:2009db,Clark:2009dc,
Barvinsky:2009ii,Barvinsky:2009jd,Lerner:2010mq,Lerner:2009na,
Giudice:2010ka,Burgess:2009,Barbon:2009ya,Burgess:2010zq,Hertzberg:2010dc,  
Buck:2010sv,Lerner:2011it,Greenwood:2012aj}. Earlier studies of
non-minimally coupled scalar fields in the context of inflation can be
also found in
Refs.~\cite{Spokoiny:1984bd,Salopek:1988qh,Fakir:1990eg}.

When the Higgs inflation model described above is rewritten in the
so-called Einstein frame, where the gravity part takes the usual
Einstein-Hilbert form, it becomes essentially non-polynomial and thus
non-renormalizable, even if the gravity part is dropped off.
Therefore, it should be understood as an effective field theory valid
only up to a certain ``cut-off'' scale. One should distinguish between
two different definitions of the ``cut-off''. Quite often  the cut-off
of the theory is understood as the energy at which  the tree level
unitarity in high-energy scattering processes is violated. A second
definition of the cut-off  is the energy associated to the onset of new
physics. As it was recently stressed in Ref.~\cite{Aydemir:2012nz},
the breaking of tree level unitarity does not imply the appearance of
new physics or extra degrees of freedom right above the corresponding
energy scale; it just signals that the perturbation theory in terms of
low-energy variables breaks down. For the case of Higgs inflation, the
tree-level scattering amplitudes {\em above the electroweak vacuum}
appear to hit the perturbative unitarity bound at energies
$\Lambda\sim M_P/\xi_h$
\cite{Burgess:2009,Barbon:2009ya,Burgess:2010zq, Hertzberg:2010dc}.
Whether the theory requires an ultraviolet completion at these
energies or simply enters into the non-perturbative strong-coupling
regime with onset of new physics  at higher energies (which could be
as large as the Planck scale) is  still an open question.
Nevertheless,  the Higgs inflation scenario is {\em self-consistent}.
As shown in Ref.~\cite{Bezrukov:2010jz} (see also
\cite{Ferrara:2010in}), the beginning of the strong coupling regime
(i.e. the cut-off scale according to the first definition which will be
used in this article) depends  on the {\em dynamical} expectation
value of the  Higgs field, which makes  the theory weakly coupled for
all the relevant energy scales in the evolution of the Universe. In
other words, the SM with a large non-minimal coupling of the Higgs
field to gravity represents a viable effective theory for the
description of  inflation, reheating, and the hot Big Bang theory.

The Higgs inflation scenario can be easily incorporated into a larger
framework, the Higgs-Dilaton model \cite{Shaposhnikov:2008xb,
GarciaBellido:2011de}. The key element of this extension is
scale-invariance (SI). No dimensional parameters such as masses are
allowed to appear in the action.  All the scales are instead induced
by the spontaneous breaking of SI. This is achieved by the
introduction of a new scalar degree of freedom, the dilaton, which
becomes the Goldstone boson  of the broken symmetry and remains
exactly massless. The coupling of the dilaton field to matter is weak
and takes place only through derivative couplings, not contradicting
therefore any 5th force experimental bounds \cite{Kapner:2006si}.

Although the dilatation symmetry described above forbids the
introduction of a cosmological constant term, the ever-present
cosmological constant problem reappears associated to the fine-tuning
of the  dilaton self-interaction \cite{Shaposhnikov:2008xb}. However,
if  the dilaton self-coupling $\beta$ is chosen to be zero (or
required to vanish due to some yet unknown reason), a slight
modification of general relativity (GR), known as Unimodular Gravity
(UG), provides a dynamical dark energy  (DE) stage in good agreement
with observations. The scale-invariant  UG gives rise to a
``run-away'' potential for the dilaton \cite{Shaposhnikov:2008xb},
which plays the role of a quintessence field. The strength of such a
potential is determined by an integration constant $\Lambda_0$ that appears in the
Einstein equations of motion due to the unimodular constraint $\hat g
= -1$ on the metric determinant.  The common origin of the
inflationary and DE dominated stages in Higgs-Dilaton inflation 
allowed to derive extra bounds on the initial inflationary
conditions\footnote{The fine-tuning needed  to reproduce the present
dark energy abundance is transferred into the initial inflationary
conditions  for the fields at the beginning of inflation.}, as well as
potentially testable relations between the early and late Universe
observables  \cite{GarciaBellido:2011de}. 

Some of the properties of the Higgs-Dilaton model described above were
previously noted in the literature. The first attempt to formulate a
viable SI theory non-minimally coupled to gravity  was done by Fujii
in Ref.~\cite{Fujii:1982ms}, although without establishing any 
connection to the SM Higgs. The role of dilatation symmetry in
cosmology  was  first considered by Wetterich in Refs.~\cite{Wetterich:1987fk,Wetterich:1987fm}. In these seminal papers, the
dynamical dark energy, associated with the  dilaton field, appears as
a consequence of the dilatation anomaly and is related to the 
breaking of SI by quantum effects. The present paper has a number of
formal analogies and similarities regarding the cosmological
consequences for the late Universe with  Refs.~\cite{Wetterich:1987fk,Wetterich:1987fm}. At the same time, our
approach to the source of dark energy is different from the one
adopted in Refs.~\cite{Wetterich:1987fk,Wetterich:1987fm}, as we
assume that SI is an exact (but spontaneously broken) symmetry at the
quantum level, leading therefore to a  massless dilaton. In Ref.~\cite{Wetterich:1987fk,Wetterich:1987fm}, both the cases of exact and
explicitly broken dilatation symmetry were considered. Our theory with
exact dilatation symmetry is different from that of~\cite{Wetterich:1987fk,Wetterich:1987fm} in two essential aspects.
First, in our work the Higgs field of the SM has non-minimal coupling
to gravity (it is absent in
Ref.~\cite{Wetterich:1987fk,Wetterich:1987fm}), which is important for
the early Universe and leads to Higgs inflation. Second, the
unimodular character of gravity (as opposed to standard general
relativity used in~\cite{Wetterich:1987fk,Wetterich:1987fm}) leads to
an automatic and very particular type of dilatation symmetry  breaking, 
which results in dynamical dark energy due to the dilaton field
(absent in~\cite{Wetterich:1987fk,Wetterich:1987fm} for the case of
exact scale invariance).
 
Our purpose here is to study, following the approach of
Ref.~\cite{Bezrukov:2010jz}, the self-consistency of the Higgs-Dilaton
model by adopting an effective field theory point of view. We will
estimate the field-dependent cut-offs associated to the different
interactions among scalars fields, gravity, vector bosons and
fermions. We will identify the lowest cut-off as a function of the
background fields and show that its value  is higher than the typical
energy scales describing the Universe during its different epochs. The
issue concerning quantum corrections generated by the loop expansion
is also addressed.  Since the model is non-renormalizable, an infinite
number of counter-terms must be added in order to absorb the
divergences. It is important to stress at this point that, in the lack of a quantum theory for
gravity, the details of the regularization scheme to be used  cannot be univocally fixed. This means that the predictions
 of the model will be sensitive to the assumptions about the UV-completion of the theory 
(corresponding to different regularization prescriptions). We will adopt a ``minimal setup" that keeps intact the
exact and approximative symmetries of the classical action and does
not introduce any extra degrees of freedom.  Within this approach, 
the relations connecting the inflationary and the dark energy
domination periods hold  even in the presence of quantum corrections.

The structure of the paper is as follows. In Section \ref{sec:model}
we briefly review the Higgs-Dilaton model. In Section
\ref{sec:cut-off} we calculate the cut-off of the theory in the Jordan
frame and compare it with the other relevant energy scales in the
evolution of the Universe. In Section \ref{sec:divergences} we propose
a ``minimal setup'' which removes all the divergences and discuss the
sensitivity of the cosmological observables to radiative corrections.
Section \ref{sec:conclusions} contains the conclusions.

\section{Higgs-Dilaton cosmology}
\label{sec:model}
We start by reviewing the main results of Refs.~\cite{Shaposhnikov:2008xb,GarciaBellido:2011de}, where the
Higgs-Dilaton model was proposed and studied in detail. The two main
ingredients of the theory are outlined below. The first one is the 
invariance of the SM action under global scale transformations,
which leads to the absence of any dimensional parameters or scales. Denoting 
by $\Phi(x)$ the field content of the theory in a metric $g_{\mu\nu}(x)$, these
 trasformations can be written as\footnote{For
a theory invariant under all diffeomorphisms, this is equivalent to 
\begin{equation*}
g_{\mu\nu}(x) \rightarrow \sigma^{-2}g_{\mu\nu}(x) \ , 
\ \ \ \Phi(x)\rightarrow\sigma^{d_\Phi}\Phi(x) \ .
\end{equation*}
}
\be{global}
g_{\mu\nu}(x)\rightarrow g_{\mu\nu}(\sigma x) \ , 
\ \ \ \Phi(x)\rightarrow \sigma^{d_\Phi}\Phi(\sigma x) \ ,
\ee
with $\sigma^{d_{\Phi}}$ the so-called scaling dimension and $\sigma$ an arbitrary constant.
In order to achieve invariance under these transformations, we let the masses
and dimensional couplings in the theory to be dynamically induced by  a field. 
The simplest choice would be to use the SM Higgs, already present in the theory. Note however that this 
option is clearly incompatible with the experiment. As discussed in Refs.~\cite{Salopek:1988qh,CervantesCota:1995tz},
 the excitations of the Higgs field in this case
become massless and completely decoupled from the  SM
particles. 

The next simplest possibility  is to introduce a new scalar singlet under
the SM gauge group. We will refer to it as the dilaton $\chi$.  The coupling between the new 
field and the SM particles, with the exception of the Higgs boson, is
forbidden by quantum numbers. The corresponding Lagrangian is given by
\be{general-theory}
\frac{\mathscr L}{\sqrt{-g}}=
\frac{1}{2}(2\xi_h \varphi^\dagger \varphi+\xi_\chi \chi^2)R+
\mathscr L_{\text{SM}[\lambda\rightarrow0]}-
\frac{1}{2}g^{\mu\nu}\partial_\mu \chi\partial_\nu\chi-V(\varphi,\chi) \ ,
\ee
where $\varphi$ is the SM Higgs field doublet 
and $\xi_h\sim 10^3-10^5, \ \xi_\chi\sim 10^{-3}, $ are respectively the
non-minimal couplings of the Higgs and dilaton fields to gravity \cite{GarciaBellido:2011de}. 
The term $\mathscr
L_{\text{SM}[\lambda\rightarrow0]}$ is the SM Lagrangian without the
Higgs potential, which in the present scale-invariant theory becomes
\be{general-potential}
V(\varphi,\chi)=\lambda\left(\varphi^\dagger \varphi-
\frac{\alpha}{2\lambda}\chi^2 \right)^2 +\beta \chi^4 \ ,
\ee
with $\lambda$ the self-coupling of the Higgs field. 

In order for this
theory to be phenomenologically viable, we demand the existence of
a symmetry-breaking ground state with non-vanishing background expectation
value for both\footnote{If $\bar\chi=0$
the Higgs field is massless, and if $\bar h=0$ there is no electroweak
symmetry breaking.}  the dilaton ($\bar\chi$) and the Higgs field in the unitary gauge
 ($\bar h$). This is given by
\be{ground-states}
\bar h^2=\frac{\alpha}{\lambda}\bar\chi^2 +\frac{\xi_h}{\lambda}R \ , 
\ \ \ \text{with} \ \ \  
R= \frac{4\beta\lambda}{\lambda\xi_\chi+\alpha\xi_h}\bar\chi^2 \,.
\ee
 All the physical scales are proportional to the non-zero background value of the
dilaton field. For instance, the SM Higgs mass is given by 
\be{Higgs-mass}
m_H^2=2\alpha M_P^2\frac{(1+6\xi_\chi)+
\frac{\alpha}{\lambda}(1+6\xi_h)}
{(1+6\xi_\chi)\xi_\chi+\frac{\alpha}{\lambda}(1+6\xi_h)\xi_h}
+\mathcal O(\beta) \ ,
\ee
with $M_P^2\equiv \xi_h \bar h^2+\xi_\chi \bar\chi^2\propto \bar\chi^2$
the effective Planck scale in the Jordan frame. The same happens with the effective cosmological 
constant
\be{cosm-const}
\Lambda=\frac{1}{4}M_P^2R=\frac{\beta M_P^4}
{(\xi_\chi+\frac{\alpha}{\lambda}\xi_h)^2+
4\frac{\beta}{\lambda}\xi_h^2} \ ,
\ee
which depending on the value of the dilaton self-coupling
$\beta$, gives rise to a
flat ($\beta=0$), deSitter ($\beta>0$) or anti-deSitter ($\beta <0$)
spacetime.
It is important to notice however that physical observables, corresponding to
dimensionless ratios between scales or masses, are independent of the
particular value of the background field $\bar\chi$. In order to reproduce the 
 ratio between the different energy scales, the parameters of the model 
must be properly fine-tuned. As shown in Eq.~\eqref{Higgs-mass}, the difference between the
electroweak and the Planck scale is encoded in the parameter\footnote{Note that the alternative choice 
$\xi_h \ggg 1$ is not compatible with CMB observations,
 cf. Eq.~\eqref{power} and Fig.\ref{fig:nsxichiRG}.} $\alpha\sim 10^{-35}\lll 1$. Similarly, the hierarchy
 between the cosmological constant and the electroweak
scale, cf. Eq.~\eqref{cosm-const}, implies  $\beta\lll\alpha$. The smallness
 of these parameters, together with the tiny value of the non-minimal coupling
$\xi_\chi$, gives rise to an approximate shift symmetry for the dilaton field at the 
classical level, $\chi\to\chi+\text{const}$. As we will show in Section \ref{sec:divergences}, this fact 
will will have important consequences for the analysis of the quantum effects.

The second ingredient of the Higgs-Dilaton cosmological model is the replacement of GR by Unimodular Gravity, which is just a 
particular case of the set of theories invariant under transverse diffeomorphisms. These
theories generically contain an extra scalar degree of freedom on top
of the massless graviton (for a general discussion see for instance Ref.~\cite{Blas:2011ac} and references therein). In UG
the number of dynamical components of the metric is effectively reduced to the standard value by 
requiring the metric determinant $\hat g$  to take some fixed constant value, conventionally
$\vert \hat g\vert = 1$. As shown in Ref.~\cite{Shaposhnikov:2008xb}, the equations of motion of a theory subject to
that constraint 
\be{unimod}
\mathscr L_{\text{UG}}= \mathscr L[\hat g_{\mu\nu},
\partial \hat g_{\mu\nu},\Phi,\partial\Phi] \ ,
\ee
coincide with those obtained from a diffeomorphism invariant theory 
 with modified action
\be{diff-inv}
\frac{\mathscr L}{\sqrt{-g}}= 
\mathscr L[g_{\mu\nu},\partial g_{\mu\nu},
\Phi,\partial\Phi] +\Lambda_0 \ .
\ee
Note that, from the point of view of UG, the parameter $ \Lambda_0$ is just a
conserved quantity associated to the unimodular constraint and it
should not be understood as a cosmological constant.

Since the two formulations are completely equivalent\footnote{As usual, there are some 
subtleties related to the quantum formulation of (unimodular) gravity. However, these will not play any role in the further 
developments. The interested reader is referred to the discussion in Ref.~\cite{Blas:2011ac} and references therein.}, we will stick to the diffeomorphism 
invariant language. Expressing the theory resulting from the
combination of the above ideas in the unitary gauge
$\varphi^T=(0,h/\sqrt{2})$ we get
\be{jord-theory}
\frac{\mathscr L}{\sqrt{-g}}=\frac{1}{2}(\xi_h h^2+\xi_\chi \chi^2)R-
\frac{1}{2}(\partial \chi)^2-\frac{1}{2}(\partial h)^2-U(h,\chi) \ ,
\ee
where the potential includes now the UG integration constant $\Lambda_0$
\be{jord-pot}
U(h,\chi)\equiv V(h,\chi)+\Lambda_0=\frac{\lambda}{4}
\left(h^2-\frac{\alpha}{\lambda}\chi^2 \right)^2+\beta \chi^4+\Lambda_0 \ .
\ee
Notice that the Lagrangian given by Eqs.~\eqref{jord-theory} and~\eqref{jord-pot} bears a clear
resemblance with the models  studied in
Ref.~\cite{Wetterich:1987fk,Wetterich:1987fm}. In particular, it
coincides (up to the non-minimal coupling of the Higgs field to
gravity) with the Brans-Dicke theory with cosmological constant studied in~\cite{Wetterich:1987fk}. However, the interpretation  of the $\Lambda_0$ term is  different.  In our case this constant is not a 
fundamental parameter associated with the anomalous breaking of SI
\cite{Wetterich:1987fm}, but an automatic result of UG.

The phenomenological consequences of  Eq.~\eqref{jord-theory} are more 
easily discussed in the Einstein frame. Let us then perform a conformal
redefinition of the metric $\tilde g_{\mu\nu}=\Omega^2g_{\mu\nu}$
with conformal factor  $\Omega^2=M_P^{-2}(\xi_h
h^2+\xi_\chi \chi^2)$. Using the standard relations  \cite{Fujii:2003pa}
\be{confor2}
\sqrt{-g}=\Omega^{-4}\sqrt{\tilde g} \ \ \ \ \  
\text{and} \ \ \ \ \  R=\Omega^{2}\left(\tilde R+
6\tilde\square\log\Omega-6\tilde g^{\mu\nu}\partial_\mu
\log\Omega \ \partial_\nu \log\Omega\right) \ ,
\ee
we get
\be{einst-theory}
\frac{\mathscr L}{\sqrt{-\tilde g}}=
\frac{M_P^2}{2}\tilde R-\frac{1}{2}\tilde K(h,\chi)-\tilde U(h,\chi) \ ,
\ee
where
\be{einst-pot}
\tilde U(h,\chi)\equiv\frac{U(h,\chi)}{\Omega^4}\equiv 
\frac{M_P^4}{(\xi_\chi \chi^2+\xi_h h^2)^2}
\left[ \frac{\lambda}{4}\left(h^2-\frac{\alpha}{\lambda}\chi^2 \right)^2+
\beta \chi^4	+\Lambda_0\right] \ ,
\ee
is the potential \eqref{jord-pot} in the new frame.
The non-canonical kinetic term in Eq.~\eqref{einst-theory} can be written as 
\be{kinetic}
\tilde K(h,\chi) = \kappa^E_{ij}
\tilde g^{\mu\nu}\partial_\mu\Phi^i\partial_\nu\Phi^j \ , 
\ee
where the quantity
\be{kmetric} 
\kappa^E_{ij}\equiv\frac{1}{\Omega^2}\left(\delta_{ij}+
\frac{3}{2}M_P^2\frac{\partial_i\Omega^2\partial_j\Omega^2}
{\Omega^2} \right) \ 
\ee
can be interpreted as the metric in the two-dimensional field space $(\Phi^1, \Phi^2)=(h,\chi)$ in the Einstein-frame. Note that, unlike the simplest Higgs inflationary
scenario  \cite{Bezrukov:2007ep},  Eq.~\eqref{kinetic} cannot be recast in canonical form by field redefinitions. In fact,  the Gaussian curvature
 associated to \eqref{kmetric} does not identically vanish unless $\xi_h=\xi_\chi$, which, as shown
  in Ref.~\cite{GarciaBellido:2011de}, is not consistent with observations. Nevertheless, it is possible to write the 
  kinetic term in a quite simple diagonal form. As
shown in Ref.~\cite{GarciaBellido:2011de}, the whole inflationary period
takes place inside a field space domain in which the contribution
of the integration constant $\Lambda_0$ is completely negligible. We will refer to this domain as the
``scale invariant region'' and  assume that it is maintained even when the
radiative corrections are taken  into account (cf. Section \ref{sec:divergences}). In this case, the dilatational 
Noether's current in the slow-roll approximation, $(1+6\xi_h)h^2+(1+6\xi_\chi)\chi^2$, 
 is approximately conserved, which suggests the definition of the set of variables
\be{var-rho-phi}
\rho= \frac{M_P}{2}\log\left[\frac{
(1+6\xi_h)h^2+(1+6\xi_\chi)\chi^2}{M_P^2} \right] \ ,\ \ \  
\tan\theta=\sqrt{\frac{1+6\xi_h}{1+6\xi_\chi}}\frac{h}{\chi} \ .
\ee
The physical interpretation of these variables is straightforward.
They are simply adequately rescaled polar variables in the $(h,\chi)$
plane. Expressed in terms of $\rho$ and $\theta$, the kinetic term \eqref{kinetic}
 turns out to be 
\be{polar-kin}
\tilde K= \left( \frac{1+6\xi_h}{\xi_h}\right)\frac{1}{\sin^2\theta+\varsigma\cos^2\theta}
(\partial \rho)^2\ +\frac{M_P^2 \ \varsigma}{\xi_\chi}
\frac{\tan^2\theta+\eta}{\cos^2\theta(\tan^2\theta+\varsigma)^2}(\partial \theta)^2 \ ,
\ee
with
\be{var-defs1}
\eta=\frac{\xi_\chi}{\xi_h} \ \
\ \text{and} \ \ \  \varsigma=\frac{(1+6\xi_h)\xi_\chi}{(1+6\xi_\chi)\xi_h} \ .
\ee 
The potential \eqref{einst-pot} is naturally divided into a scale-invariant part, depending only on the $\theta$ field, and
 a scale-breaking part, proportional to $\Lambda_0$ and depending on both $\theta$ and $\rho$. These are respectively given by 
\be{polar-pot}
\tilde U(\theta)= \frac{\lambda M_P^4}{4\xi_h^2}
\left(\frac{\sin^2\theta}
{\sin^2\theta+\varsigma\cos^2\theta}\right)^2 \ , \hspace{3mm}
 \tilde U_{\Lambda_0}(\rho,\theta)=\Lambda_0 \left( \frac{1+6\xi_h}{\xi_h}\right)^2
\frac{e^{-4\rho/M_P}}{(\sin^2\theta+\varsigma\cos^2\theta)^2} \ , 
\ee
where we have safely neglected the contribution of $\alpha$ and $\beta$ in Eq.~\eqref{einst-pot}.
Note that the non-minimal couplings of the fields to gravity with $\Lambda_0>0$ naturally generate a 
``run-away'' potential for the physical dilaton, similar  to those considered in the
 pioneering works on quintessence \cite{Wetterich:1987fk,Wetterich:1987fm,Ratra:1987rm}.

The inflationary period of the expansion of the Universe 
takes place for field values $ \xi_h h^2 \gg \xi_\chi\chi^2$.   
From the definition of the angular variable $\theta$ in Eq.~\eqref{var-rho-phi}, this 
 corresponds to\footnote{Strictly speaking, the condition $\tan^2\theta\gg\eta$ holds beyond the inflationary
  region $ \xi_h h^2 \gg \xi_\chi\chi^2$ and includes also the reheating stage.}  $\tan^2\theta\gg\eta$. 
In that limit,
we can neglect the $\eta$ term in the kinetic term \eqref{polar-kin} and perform an extra
field redefinition
\be{var-r-theta}
r= \gamma^{-1}\rho\ \ \ \ \text{and} \ \ \   
\vert\phi'\vert=\phi_0- \frac{M_P}{a}\tanh^{-1}\left[\sqrt{1-\varsigma}\cos\theta \ \right]  \ , 
\ee
where 
\be{var-defs2}
\gamma= \sqrt{\frac{\xi_\chi}{1+6\xi_\chi}} 
\ \ \ \text{and} \ \ \ a=\sqrt{\frac{\xi_\chi(1-\varsigma)}{\varsigma}} \ .
\ee
The variable $\phi'$ is periodic and defined in the
compact interval $\phi'
\in\left[-\phi_0,\phi_0\right]$, with $\phi_0=
M_P/a \ \tanh^{-1}\left[\sqrt{1-\varsigma}\ \right] $  the value of the 
field at the beginning of inflation. 
In terms of these variables the Lagrangian \eqref{einst-theory} takes a
very simple form\footnote{Note that the definition of the angular variable 
$\phi$ used in this work is slightly different from that appearing in 
Ref.~\cite{GarciaBellido:2011de}. The new parametrization makes explicit the
 symmetry of the potential and shifts its minimum to make it coincide with that
  in Higgs-inflation.}
\be{angul-theory}
\frac{\mathscr L}{\sqrt{-\tilde g}}=\frac{M_P^2}{2}\tilde R -
\frac{\varsigma\cosh^2[a\phi/M_P]}{2}(\partial r)^2- 
\frac{1}{2}(\partial \phi)^2-\tilde U(\phi)-
\tilde U_{\Lambda_0}(r,\phi) \ ,
\ee
with $\phi= \phi_0-\vert\phi'\vert$. 
\begin{figure}
\centering
\includegraphics[scale=0.9]{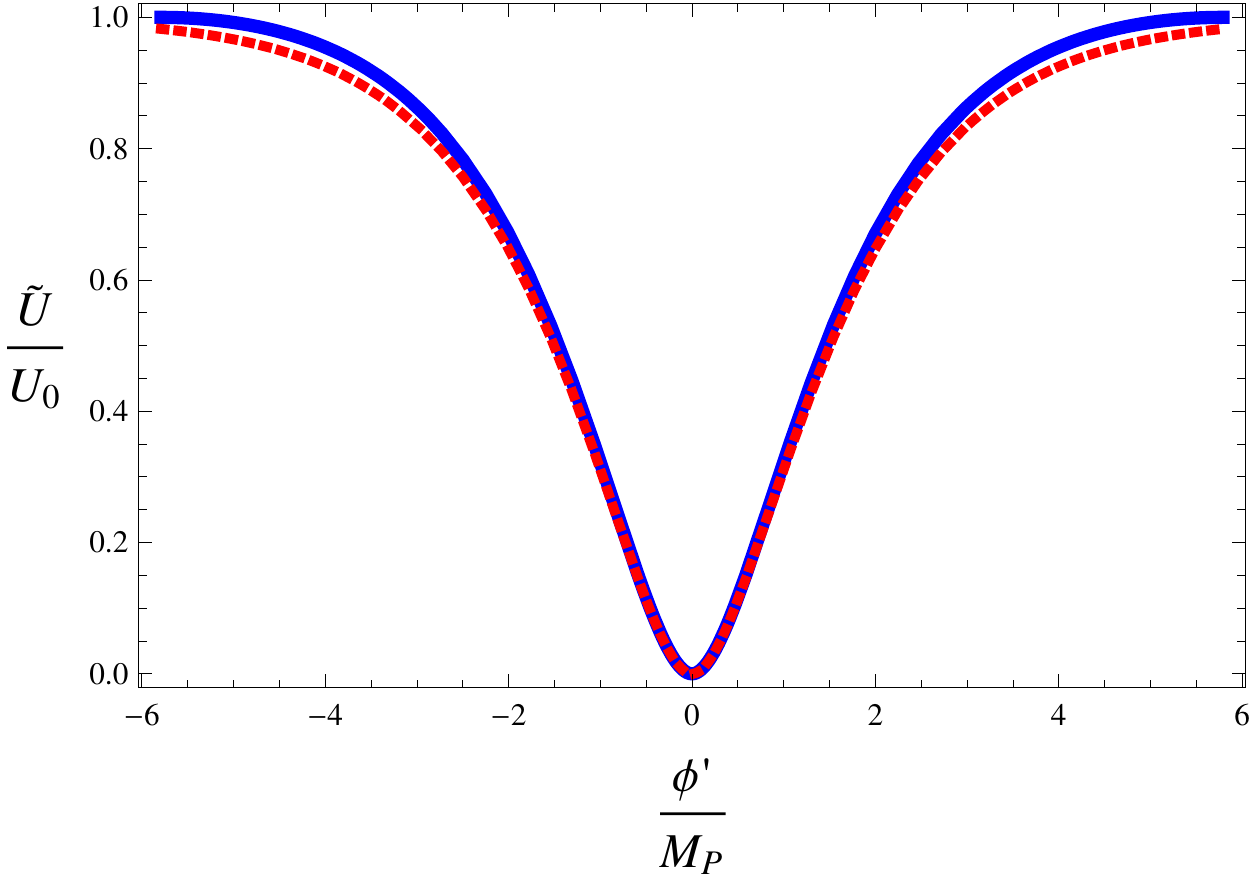}
\caption{Comparison between the Higgs-Dilaton inflationary potential
(blue continuous line) obtained from \eqref{defin-pot} in the
scale-invariant region  and the corresponding one  for the Higgs
Inflation model (red dotted line). The amplitudes are normalized
to the asymptotic value $U_0=\frac{\lambda M_P^4}{4\xi_h^2}$.} 
\label{fig:pot-comparison}
\end{figure}
The potential \eqref{polar-pot} becomes 
\be{defin-pot}
\tilde U(\phi)=\frac{\lambda M_P^4}{4\xi_h^2 (1-\varsigma)^2}
\left(1- \varsigma\cosh^2[a\phi/M_P]\right)^2 \ ,\hspace{5mm}
\tilde U_{\Lambda_0}(r,\phi)=\frac{\Lambda_0}{\gamma^4}\varsigma^2
\cosh^4[a\phi/M_P]e^{-4\gamma r/M_P} \ ,
\ee
whose scale-invariant part $\tilde U(\phi)$ resembles
 the potential of the simplest Higgs inflationary scenario \cite{Bezrukov:2007ep}, cf. Fig.~\ref{fig:pot-comparison}.  
 The analytical expressions for the amplitude and the spectral tilt of scalar perturbations at 
order $\mathcal O(\xi_\chi,1/\xi_h,1/N^*)$
 can be easily calculated to obtain  \cite{GarciaBellido:2011de}
 \be{power}
P_\zeta(k_0)\simeq \frac{\lambda\sinh^2[4\xi_\chi N^*]}
{1152\pi^2\xi_\chi^2\xi_h^2} \ ,\hspace{10mm}
n_s(k_0)\simeq 1 -8\xi_\chi\coth(4\xi_\chi N^* ) \ ,
\ee
where $N^*$ denotes the number of e-folds between the moment at which 
the pivot scale $k_0/a_0=0.002 \ \text{Mpc}^{-1}$ exited the horizon 
and the end of inflation. Note that for  $1< 4\xi_\chi
N^*\ll 4N^*$, the expression for the tilt simplifies and becomes linear in $\xi_\chi$
\be{as-tilt}
n_s(k_0)\simeq 1-8\xi_\chi \ .
\ee

An interesting cosmological phenomenology arises with the peculiar
choice\footnote{Some arguments in favour of the $\beta=0$
case can be found in Ref.~\cite{Shaposhnikov:2008xi,
GarciaBellido:2011de,Blas:2011ac}.} $\beta=0$. In this case, the DE
dominated period in the late Universe depends only on  the dilaton
field $\rho$, which give rise to an intriguing relation between the inflationary and DE domination 
periods. Let us start by noticing that around the minimum
 of the potential the value of $\theta$ is very close to zero. In that limit, $\tan^2\theta\ll\eta$,  which prevents
 the use of the field redefinition \eqref{var-r-theta}. The appropriate redefinitions needed to diagonalize the kinetic term \eqref{polar-kin}
  in this case turn out to be
\be{low-en}
r=\gamma^{-1}\rho \ \ \ \text{and} \ \ \ \phi'\simeq  
\frac{M_P}{\sqrt{\xi_h \varsigma}}\theta \ .
\ee
Using Eqs.~\eqref{polar-kin} and \eqref{polar-pot} it is straightforward 
to show that the part of the theory associated to the Higgs field $\phi$
 simplifies to the SM one. The resulting scale-invariance breaking potential for
the dilaton is still of the ``run-away'' type
\be{darkener-pot}
\tilde U_{\Lambda_0}(r)=
\frac{\Lambda_0}{\gamma^4}e^{-4\gamma r/M_P} \ ,
\ee
 making it suitable for playing the role of quintessence. 
Let us assume that 
$\tilde U_{\Lambda_0}$ is negligible during the radiation
and matter dominated stages but responsible for the present
accelerated expansion of the Universe. In that case, it is possible to
write the following relation between the equation of state parameter
$\omega_r$ of the $r$ field and its  relative abundance $\Omega_r$
\cite{Scherrer:2007pu}
\be{eosp-abund}
1+\omega_r=\frac{16\gamma^2}{3}\left[ \frac{1}{\sqrt{\Omega_r}}-
\frac{1}{2}\left(\frac{1}{\Omega_r}-1\right)\log\frac{1+
\sqrt{\Omega_r}}{1-\sqrt{\Omega_r}}\right]^2 \ . 
\ee
For the present DE density $\Omega_\text{DE}=\Omega_r\simeq 0.74$, the above expression yields
\be{reduc}
1+\omega_\text{DE}=\frac{8}{3}\frac{\xi_\chi}{1+6\xi_\chi} \ .
\ee
Comparing Eqs.~\eqref{as-tilt} and \eqref{reduc}, it follows that the
deviation of the scalar tilt $n_s$  from the scale-invariant one is
proportional to the deviation of the DE equation of state from a cosmological
constant\footnote{Outside this region of parameter space, the relation
connecting $n_s \ \text{to} \ \omega_\text{DE}$ is somehow more
complicated 
\be{functional-rel-compl}\nonumber
n_s-1 \simeq -\frac{12(1+\omega_\text{DE})}
{4-9(1+\omega_\text{DE})}\coth\left[\frac{6N^*(1+
\omega_\text{DE})}{4-9(1+\omega_\text{DE})} \right] \ .
\ee} \cite{GarciaBellido:2011de}
\be{functional-rel}
n_s-1 \simeq -3(1+\omega_\text{DE}), \ \ \ \text{for} \ \ \ 
  \frac{2}{3N^*} < 1+\omega_\text{DE}\ll 1 \ . 
\ee
The above condition is a non-trivial
prediction of Higgs-Dilaton cosmology, relating two a priori
completely independent periods in the history of the Universe. This
has interesting consequences from an observational point of
view\footnote{Similar consistency relations relating the rate of
change of the equation of state parameter $w(a) = w_0 + w_a (1 - a)$
with the logarithmic running of the scalar tilt can be also derived,
cf. Ref.~\cite{GarciaBellido:2011de}. The practical relevance of those
consistence conditions is however much more limited than that of 
Eq.~\eqref{functional-rel}, given the small value of the running of the
scalar tilt in Higgs-driven scenarios.} and makes the Higgs-Dilaton
scenario rather unique. We will be back to this point
 in Section \ref{sec:divergences}, where  we
will show that the consistency relation \eqref{functional-rel-compl} still holds even in  the
 presence of quantum corrections computed within the ``minimal setup''. 

\section{The dynamical cut-off scale}
\label{sec:cut-off}
Following Ref.~\cite{Bezrukov:2010jz}, we now turn to the determination of
the energy domain where the Higgs-Dilaton model can be considered as a
predictive effective field theory. This domain is bounded from above
by the field-dependent cut-off $\Lambda(\Phi)$, i.e. the energy where
perturbative tree-level unitarity is violated \cite{Cornwall:1974km}.
At energies above that scale, the theory becomes strongly-coupled and
the standard perturbative methods fail. In order to determine this
 (background dependent) energy scale, two
related methods, listed below, can be used.
\begin{enumerate}
\item[(1)] Expand the generic fields of the theory around their background
values 
\be{split}
\Phi(\mathbf x,t)=\bar\Phi+\delta\Phi(\mathbf x,t) \ ,
\ee 
such that all kind of higher-dimensional non-renormalizable operators
\be{non-renorm-op}
c_n\frac{\mathcal O_n(\delta\Phi)}{[\Lambda(\bar\Phi)]^{n-4}} \ ,
\ee
with $c_n\sim\mathcal O(1)$ appear in the resulting action. These
operators are suppressed by appropriate powers of the field-dependent
coefficient $\Lambda(\bar\Phi)$, which can be identified as the
cut-off of the theory.  This procedure gives us only a lower estimate
of the cut-off, since it does not take into account the possible
cancelations that might occur between the different scattering diagrams.
\item[(2)] Calculate at which energy each of the N-particle scattering
amplitudes hit the unitarity bound. The cut-off will then be the
lowest of these scales. 
\end{enumerate}

In what follows we will apply these two methods to determine the
effective cut-off of the theory. We will start by applying the
method $(1)$ to compute the cut-off associated with the
gravitational and scalar interactions. The cut-off associated to the
gauge and fermionic sectors will be obtained via the method
$(2)$.

\subsection{Cut-off in the scalar-gravity sector}

We choose to work in the original Jordan frame where the Higgs and
dilaton fields are non-minimally coupled to gravity\footnote{A similar study in the Einstein frame can be found in the Appendix A.}. Expanding these 
fields around a static background\footnote{Note that, in comparison with 
the analysis performed in Ref.~\cite{Lerner:2011it} for generalized Higgs inflationary models, both the dilaton and
 the Higgs field acquire a non-zero background expectation value, cf. Section \ref{sec:model}. As we will see below, 
 this will give rise to a much richer cut-off structure.}
 \be{expansion}
g_{\mu\nu}=\bar g_{\mu\nu}+\delta g_{\mu\nu} \ , \hspace{.5cm}
\chi=\bar\chi+\delta\chi \ , \hspace{.5cm} h=\bar h+\delta h \ ,
\ee
we obtain the following kinetic term for the quadratic Lagrangian of
the gravity and scalar sectors
\be{quadratic}
\begin{aligned}
\mathscr K_2^{\text{G+S}}=\frac{\xi_\chi \bar \chi^2 +\xi_h\bar
h^2}{8}\left(\delta g^{\mu\nu}\square \delta g_{\mu\nu}+2\partial_\nu
\delta g^{\mu\nu}\partial^\rho\delta g_{\mu\rho}-2\partial_\nu \delta
g^{\mu\nu}\partial_{\mu}\delta g -\delta g\square \delta g\right)&\\
-\frac{1}{2}(\partial \delta \chi)^2 -\frac{1}{2}(\partial \delta h)^2
+(\xi_\chi \bar\chi\delta\chi +\xi_h \bar h \delta
h)(\partial_\lambda\partial_\rho\delta g^{\lambda\rho}-\square\delta
g) \ .&
\end{aligned}
\ee
The leading higher-order non-renormalizable operators obtained in this
way are given by
\be{interact}
 \xi_\chi(\delta \chi)^2\square \delta g \ , \ \ \  \xi_h(\delta
h)^2\square \delta g \ .
\ee
Note that these operators are written in terms of quantum excitations
with non-diagonal kinetic terms. In order to properly identify the
cut-off of the theory, we should determine the normal modes that
diagonalize the quadratic Lagrangian \eqref{quadratic}. After doing that,
and using the equations of motion to eliminate artificial degrees of
freedom, we find that the metric perturbations depend on the scalar
fields perturbations, a fact that is implicit in the Lagrangian
\eqref{quadratic}. The gravitational part of  the above action 
can be recast into canonical form in terms of a new
metric perturbation $\delta\hat g_{\mu\nu}$ given by
\be{redefinition-metric}
\delta\hat g_{\mu\nu}=\frac{1}{\sqrt{\xi_\chi\bar \chi^2 +\xi_h \bar
h^2}}\left[(\xi_\chi\bar \chi^2 +\xi_h \bar h^2)\delta  g_{\mu\nu}
+2\bar g_{\mu\nu}(\xi_\chi\bar\chi \delta\chi+\xi_h\bar h \delta h)
\right] \ .
\ee
The cut-off scale associated to  purely gravitational interactions
becomes in this way the effective Planck scale in  the Jordan frame 
\be{gravcut-off}
\Lambda_{P}^2=\xi_\chi\bar \chi^2+ \xi_h \bar h^2 \ .
\ee
The remaining non-diagonal kinetic term for the scalar perturbations $(\delta\Phi^1, \delta\Phi^2)=(\delta h,\delta \chi)$ is given
in compact matrix notation by
\be{matrixnot}
\mathscr K_2^{\text{S}}=-\frac{1}{2}\bar\kappa^J_{ij}\partial_\mu
\delta\Phi^i\partial^\mu\delta\Phi^j \ ,
\ee 
where $\bar \kappa^J_{ij}=\Omega^2 \bar \kappa^E_{ij}$ is the Jordan frame 
analogue of Eq.~\eqref{kmetric} and depends
only on the background values of the fields, i.e.
\be{kinetic-fields}
\bar\kappa^J_{ij}= \frac{1}{\xi_\chi\bar\chi^2+\xi_h\bar
h^2}\begin{pmatrix}\xi_\chi \bar\chi^2(1+6\xi_\chi) +\xi_h\bar h^2&
6\xi_\chi \bar\chi\xi_h\bar h\\ 6\xi_\chi \bar\chi\xi_h\bar h&\xi_\chi
\bar\chi^2 +\xi_h\bar h^2(1+6\xi_h)
\end{pmatrix} \ .
\ee
In order to diagonalize the above expression we make use of the
following set of variables
\be{redefinition-fields}
\begin{aligned}
&\delta\hat\chi=\sqrt{\frac{\xi_\chi\bar \chi^2(1+6\xi_\chi)
+\xi_h \bar h^2(1+6\xi_h)}{(\xi_\chi^2\bar \chi^2 +\xi_h^2 \bar
h^2)(\xi_\chi\bar \chi^2 +\xi_h \bar h^2)}}\left(\xi_\chi\bar\chi
\delta\chi+\xi_h\bar h \delta h\right) \
, \\ &\delta \hat h=\frac{1}{\sqrt{\xi_\chi^2\bar \chi^2 +\xi_h^2 \bar
h^2}}\left(-\xi_h\bar h \delta\chi+\xi_\chi\bar\chi\delta h\right) \ .
\end{aligned}
\ee
Note here that this is precisely the change of
variables (up to an appropriate rescaling with the conformal factor
$\Omega$) needed to diagonalize the kinetic terms for the scalar
perturbations in the Einstein frame. To see this, it is enough to
start from Eq.~\eqref{kinetic} and expand the fields around their
background values $\Phi^i\rightarrow \bar\Phi^i+\delta\Phi^i$. Keeping
the terms with the lowest power in the excitations, $\tilde K=
\bar\kappa^E_{ij}\partial_\mu\delta\Phi^i\partial^\mu\delta\Phi^j+\mathcal
O(\delta\Phi^3)$, it is straightforward to show that the previous
expression can be diagonalized in terms of 
\be{redefinition-fields-einstein}
\begin{aligned}
&\delta \hat\chi=\bar\Omega^{-1}\sqrt{\frac{\xi_\chi\bar \chi^2(1+6\xi_\chi) +\xi_h
\bar h^2(1+6\xi_h)}{(\xi_\chi^2\bar \chi^2 +\xi_h^2 \bar
h^2)(\xi_\chi\bar \chi^2 +\xi_h \bar h^2)}}\left(\xi_\chi\bar\chi
\delta\chi+\xi_h\bar h \delta h\right) \ , \\ 
&\delta\hat h=\bar\Omega^{-1}\frac{1}{\sqrt{\xi_\chi^2\bar \chi^2
+\xi_h^2 \bar h^2}}\left(-\xi_h\bar h
\delta\chi+\xi_\chi\bar\chi\delta h\right) \ .
\end{aligned}
\ee 
Written in terms of the canonically normalized
variables \eqref{redefinition-metric} and \eqref{redefinition-fields}
these operators read
\be{higherdimensional}
\frac{1}{\Lambda_1}(\delta\hat h)^2\square \delta \hat g \ , \ \ \ 
\frac{1}{\Lambda_2}(\delta\hat \chi)^2\square \delta \hat g \ , \ \ \ 
\frac{1}{\Lambda_3}(\delta\hat\chi)(\delta\hat h) \square \delta \hat
g \ ,
\ee
where the different cut-off scales are given by
\begin{align}
\centering
\label{cut-off1}
\Lambda_1&=\frac{\xi_\chi^2\bar \chi^2 +\xi_h^2 \bar
h^2}{\xi_\chi\xi_h\sqrt{\xi_\chi\bar \chi^2 +\xi_h \bar h^2}} \ ,\\
\label{cut-off2}
\Lambda_2&=\frac{(\xi_\chi^2\bar \chi^2 +\xi_h^2 \bar
h^2)(\xi_\chi\bar \chi^2(1+6\xi_\chi) +\xi_h \bar
h^2(1+6\xi_h))}{(\xi_\chi^3\bar \chi^2 +\xi_h^3 \bar
h^2)\sqrt{\xi_\chi\bar \chi^2 +\xi_h \bar h^2}} \ , \\
\label{cut-off3}
\Lambda_3&=\frac{(\xi_\chi^2\bar \chi^2 +\xi_h^2 \bar
h^2)(\xi_\chi\bar \chi^2(1+6\xi_\chi) +\xi_h \bar
h^2(1+6\xi_h))}{\xi_\chi\bar \chi\xi_h\bar h\left\vert \xi_h
-\xi_\chi\right\vert\sqrt{\xi_\chi\bar \chi^2 +\xi_h \bar h^2}} \ .
\end{align}
The effective cut-off of the scalar theory at a given value of the background
fields will be the lowest of the previous scales. We will be back to this point in 
Section \ref{sec:comparison}.

\subsection{Cut-off in the gauge and fermionic sectors}\label{sec:gauge-cut-off}
Let us now move to the cut-off associated with the gauge sector. Since
we are working in the unitary gauge for the SM fields, it
is sufficient to look at the tree-level scattering of non-abelian
vector fields with longitudinal polarization. It is well known that in
the SM the ``good'' high energy behaviour of these processes is 
the result of cancellations that occur when we take into account the
interactions of the gauge bosons with the excitations $\delta h$ of
the Higgs field\footnote{In the absence of the Higgs field, the scattering
amplitudes grow as the square of the center-of-mass energy, due to
the momenta dependence of the longitudinal vectors $\sim q^\mu/m_W$.} 
\cite{Lee:1977yc,Lee:1977eg}. 

In our case, even though purely gauge interactions remain unchanged,
the graphs involving the Higgs field excitations are modified due to
the non-canonical kinetic term. This changes the pattern of  the
cancellations that occur in the standard Higgs mechanism, altering
therefore the asymptotic behaviour of these processes. As a result, the
energy scale where this part of the theory becomes strongly coupled becomes lower. 

To illustrate how this happens, let us consider the  $W_L
W_L\rightarrow W_L W_L$ scattering in the $s-$channel. The relevant
part of the Lagrangian is 
\be{lagr-gauge}
g\, m_{W}W_\mu^+ W^{-\mu}\delta h \ , 
\ee
where $m_W\sim g \bar h$. After diagonalizing the kinetic term for the
scalar fields with the change of variables
\eqref{redefinition-fields}, the above expression becomes
\be{lagr-gauge-diag}
g'  m_W W_\mu^+ W^{-\mu}\delta \hat h + g''  m_W W_\mu^+
W^{-\mu}\delta \hat \chi \ , 
\ee
where the effective coupling constants $g' \ \text{and} \ g''$ are
given by
\be{eff-coupl}
g' = g \frac{\xi_\chi\bar\chi}{\sqrt{\xi_\chi^2\bar\chi^2+\xi_h^2 \bar
h^2}} \ , \ \ \ \ g'' = g \frac{\xi_h\bar
h}{\sqrt{\xi_\chi^2\bar\chi^2+\xi_h^2 \bar
h^2}}\sqrt{\frac{\xi_\chi\bar\chi^2+\xi_h \bar
h^2}{\xi_\chi\bar\chi^2(1+6\xi_\chi)+\xi_h\bar h^2(1+6\xi_h)}} \ .
\ee

From the requirement of tree unitarity of the $S$-matrix, it is
straightforward to show that the scattering amplitude of this
interaction hits the perturbative unitarity bound at energies
\be{gauge-cut-off}
\Lambda_G\simeq\sqrt{
\frac{\xi_\chi\bar\chi^2(1+6\xi_\chi)+\xi_h\bar
h^2(1+6\xi_h)}{6\xi_h^2}} \ . 
\ee
It is interesting to compare the previous expression with the results for the gauge cut-off of the simplest
Higgs inflationary model \cite{Bezrukov:2010jz}. In order to do that, let us consider two limiting cases: 
 the inflationary/high-energy period corresponding to field values $\xi_\chi \chi^2 \ll \xi_h h^2$ and the 
 low-energy regime at which  $\xi_\chi \chi^2 \gg \xi_h h^2$ . In these two cases, the above 
 expression simplifies to
\be{lim-gauge} 
\Lambda_G\simeq \Bigg\{ \begin{array}{cl} \bar h
&\mbox{for} \ \xi_\chi\bar\chi^2\ll \xi_h \bar h^2 \ , \\
\frac{\sqrt{\xi_\chi}\bar\chi}{\xi_h} &\mbox{for} \
\xi_\chi\bar\chi^2\gg \xi_h \bar h^2 \ , \\
\end{array} 
\ee
in agreement with the Higgs inflation model. 

To identify the cut-off of the fermionic part of the Higgs-Dilaton
model, we consider the chirality non-conserving process $\bar f
f\rightarrow W_L W_L$. This interaction receives contributions from
diagrams with $\gamma$ and $Z$ exchange  ($s-$channel) and from a
diagram with fermion exchange ($t-$channel). In the asymptotic
high-energy limit, the total amplitude of these graphs grows linearly
with the energy at the center of mass. Once again, the $s-$channel
diagram including the Higgs excitations unitarizes the associated amplitude
\cite{Chanowitz:1978uj,Chanowitz:1978mv,Appelquist:1987cf}.  Following
therefore the same steps as in the calculation of the gauge cut-off,
we find that this part of the theory enters into the strong-coupling
regime at energies
\be{ferm-cut-off}
\Lambda_F\simeq y^{-1}
\frac{\xi_\chi\bar\chi^2(1+6\xi_\chi)+\xi_h\bar
h^2(1+6\xi_h)}{6\xi_h^2 \bar h} \ , 
\ee
where $y$ is the Yukawa coupling constant. The above cut-off is higher
than that of the SM gauge interactions \eqref{gauge-cut-off}
during the whole evolution of the Universe. 

\subsection{Comparison with the energy scales in the early and late Universe}\label{sec:comparison}

In this section we compare the cut-offs found above with the
characteristic energy scales in the different periods during the
evolution of the Universe. If the typical momenta involved in the
different processes are sufficiently small, the theory will remain in
the weak coupling limit, making the Higgs-Dilaton scenario
self-consistent.

\begin{figure}
\centering
\includegraphics[scale=1]{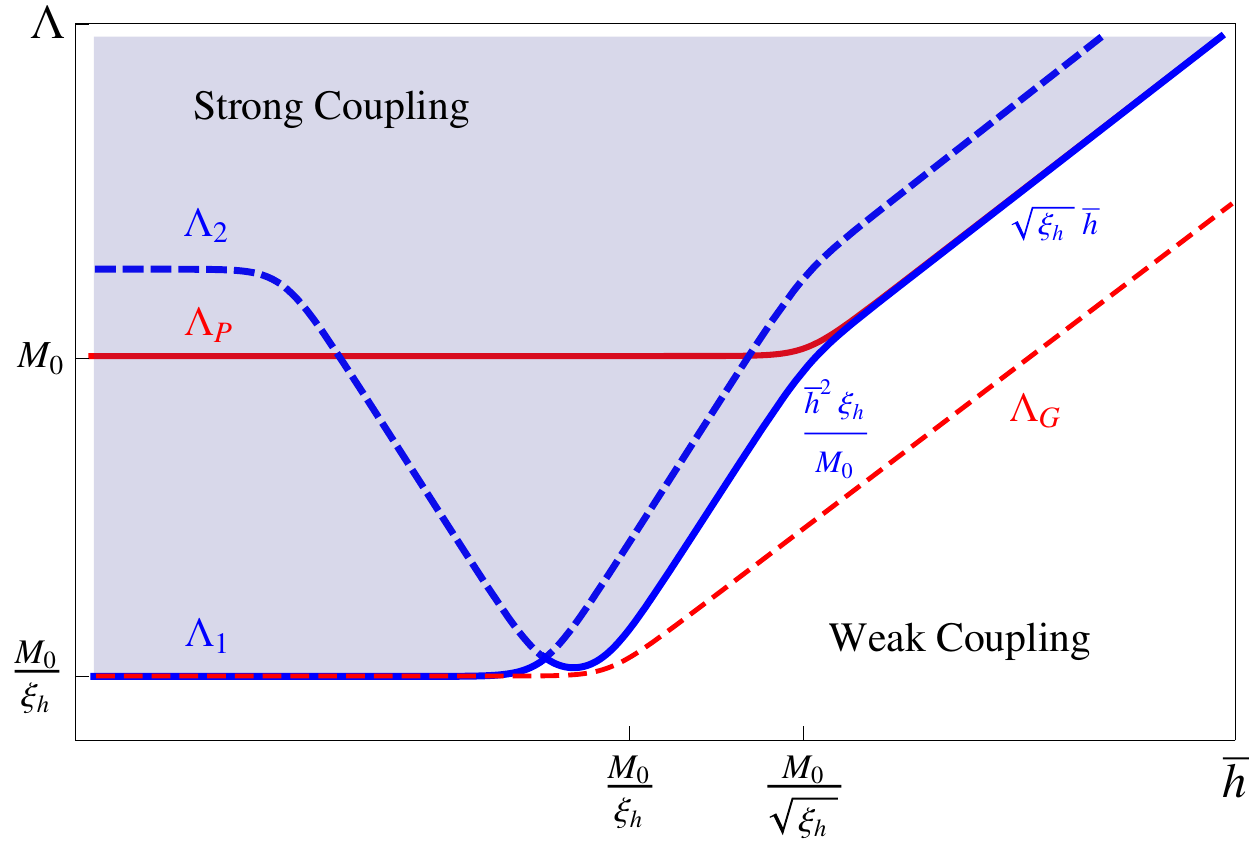}
\caption{Dependence of the different cut-off scales for a fixed value
of the dilaton field $\bar\chi$ as a function of the Higgs field $\bar
h$ in the Jordan frame. The cut-off \eqref{cut-off3} is parametrically
above the other energy scales ($\Lambda_1, \ \Lambda_2,\
\Lambda_P,\  \Lambda_G \ \text{and} \ \Lambda_F$) during the
whole history and it is therefore not included in the figure. The
effective field theory description of scalar fields is applicable for
typical energies below the thick blue solid line, which correspond to
the minimum of the scalar cut-off scales at a given field value. This
is  given by  $\Lambda_2$ and $\Lambda_1$ in the scalar sector, for
large and small Higgs values respectively. The red solid line
correspond to the gravitational cut-off \eqref{gravcut-off}, while the
red dashed one corresponds to the gauge cut-off \eqref{gauge-cut-off}.
They coincide with the effective scalar cut-off for the limiting values of the Higgs field. 
 The scale $M_0$ is defined as
$M_0=\sqrt{\xi_\chi}\bar\chi$ and corresponds to the value of the effective
Planck mass at low energies.}
\label{fig:cut-off-comparison}
\end{figure}

Let us start by considering the inflationary period, characterized by 
$\xi_h\bar h^2 \gg \xi_\chi\bar\chi^2$. As shown in Fig.~\ref{fig:cut-off-comparison}, the 
lowest cut-off in this region is the
one associated with the gauge interactions $\Lambda_G$. The
typical momenta of the scalar perturbations produced during inflation
are of the order of the Hubble parameter at that time. This quantity
can be easily estimated in the Einstein frame, where it is basically
determined by the energy stored in the inflationary potential
\eqref{defin-pot}. We obtain $\tilde H\sim \sqrt{\lambda}M_P/\xi_h$.
When transformed to the Jordan frame ($H=\Omega \tilde H$) this
quantity becomes $H\sim \sqrt{\frac{\lambda}{\xi_h}}\bar h$, which is
significantly below the cut-off scale $\Lambda_G$ in that
region. The same conclusion is obtained for the total energy density,
which turns out to be much smaller than $\Lambda_G^4$.
Moreover, the cut-off $\Lambda_G$ exceeds the masses of all
particles in the Higgs background, allowing a
self-consistent estimate of radiative corrections (cf. Section~\ref{sec:divergences}). 

After the end of inflation, the field $\phi$ starts to oscillate
around the minimum of the potential with a decreasing amplitude, due to
the expansion of the Universe and particle production. This
amplitude varies between $M_0/\sqrt{\xi_h}$ and
$M_0/\xi_h$, where $M_0=\sqrt{\xi_\chi}\bar \chi$ is
the asymptotic Planck scale in the low energy regime. As shown in Fig.~\ref{fig:pot-comparison}, the 
curvature of the Higgs-Dilaton potential
around the minimum coincides (up to ${\cal{O}}(\xi_\chi)$ 
corrections) with that of the Higgs-inflation scenario.  All
the relevant physical scales, including the effective gauge and
fermion masses, agree, up to small corrections, with those in
Higgs-inflation \cite{GarciaBellido:2012zu} . This allows us to directly apply the results of
\cite{Bezrukov:2008ut,GarciaBellido:2008ab,Bezrukov:2011sz} to the
Higgs-Dilaton scenario. According to these works, the typical momenta
of the gauge bosons produced at the minimum of the potential  in the
Einstein frame is of order $\tilde k\sim (\tilde m_A/M)^{2/3}M$, with
$\tilde m_A$ the mass of the gauge bosons in the Einstein frame and
$M=\sqrt{\lambda/3}M_P/\xi_h$ the curvature of the potential around
the minimum. After transforming to the Jordan frame we obtain $k\sim
\left(\frac{\lambda
g^4}{\xi_h}\right)^{1/6}\Lambda_G$, with $g$ the weak
coupling constant. The typical momentum of the created gauge bosons is
therefore parametrically below the gauge cut-off scale
\eqref{lim-gauge} in that region.

At the end of the reheating period, $\xi_\chi\bar\chi^2\gg \xi_h \bar h^2$, the system settles down to the
minimum of the potential $\tilde U(\phi)$, cf. Eq.~\eqref{defin-pot}. In that region the
effective Planck mass coincides with the value $M_0$. The cut-off
scale becomes  $\Lambda_1\simeq \sqrt{\xi_\chi}\bar\chi/\xi_h\simeq 
M_P/\xi_h$.
This value is much higher than the electroweak scale $m_H^2\sim
2\alpha/\xi_\chi M_P$ (cf. Eq.~\eqref{Higgs-mass}) where all the
physical processes take place. We conclude therefore that perturbative
unitarity is maintained for all the relevant processes 
during the whole evolution of the Universe.

\section{Quantum corrections}
\label{sec:divergences}

In this section we concentrate on the radiative corrections to the
inflationary potential  and on their influence on the predictions of
the model.

Our strategy is as follows. We regularize the quantum theory in such a
way that all multi-loop diagrams are finite, whereas the exact
symmetries of the chosen classical action (gauge, diffeomorphisms and
scale invariance) remain intact. Moreover, we will require the
regularization to respect the approximate  shift symmetry of the
dilaton field in the Jordan frame, cf. Section \ref{sec:model}.  Then
we add to the classical action an infinite number of counter-terms
(including the finite parts as well) which remove all the divergences
from the theory and do not spoil the exact and approximate symmetries
of the classical action. Since the theory is not renormalizable, these
counter-terms will have a different structure  from that of the
classical action.  In particular, terms that are non-analytic with
respect to the Higgs and dilaton  fields will appear
\cite{Shaposhnikov:2009nk}. They can be considered as
higher-dimensional operators, suppressed by the field-dependent
cut-offs. For consistency with the analysis made earlier in this work,
we demand these cut-offs to exceed those found in Section 
\ref{sec:cut-off}.

An example of the subtraction procedure which satisfies all the
requirements  formulated above has been constructed in 
Ref.~\cite{Shaposhnikov:2008xi} (see also earlier discussion in
\cite{Englert:1976ep}). It is based on dimensional regularization in
which the 't Hooft-Veltman normalization point $\mu$ is replaced by
some combination of the scalar fields with an appropriate dimension,
$\mu^2\to F(\chi,h)$ (we underline that we use the Jordan frame here
for all definitions). The infinite part of the counter-terms is
defined as in $\overline{MS}$ prescription, i.e. by subtracting the
pole terms in $\epsilon$, where the dimensionality of space-time is
$D=4-2\epsilon$. The finite part of the counter-terms has the same
operator structure as the infinite part, including the parametric
dependence on the coupling constants.  

Although the requirement of the structure of higher-dimensional
operators, formulated in the previous paragraphs puts important
constraints on the function $F(\chi,h)$, its precise form is not
completely determined
\cite{Shaposhnikov:2008xi,Shaposhnikov:2009nk,Codello:2012sn}, and the
physical results {\em do depend} on the choice of $F(\chi,h)$. This
somewhat mysterious fact from the point of view of uniquely defined
classical theory (\ref{general-theory}) becomes clear if we  recall
that we are dealing with a non-renormalizable theory. The quantization
of this kind of theories requires the choice of a particular classical
action together  with a set of subtraction rules. The ambiguity in the
choice of the field-dependent  normalization point $F(\chi,h)$ simply
reflects our ignorance about the proper set of rules. Different
subtractions prescriptions applied to  the same classical action do
produce unequal results. Sometimes this ambiguity is formulated as a
dependence of quantum theory on the choice of  conformally related
frames in scalar-tensor theories \cite{Flanagan:2004bz}. The use of
the {\em same} quantization rules in different frames would lead to
quantum theories with different choices of  $F(\chi,h)$.

Among the many possibilities, the simplest  and most natural choice 
is to  identify the normalization point in the Jordan frame with the
gravitational cut-off \eqref{gravcut-off},
\begin{equation}
\label{prescription1J}
\mu_I^2 \propto \xi_\chi
\chi^2 + \xi_h h^2,
\end{equation}
which corresponds to the scale-invariant
prescription of Ref.~\cite{Shaposhnikov:2008xi}. In the Einstein frame
the previous choice becomes standard (field-independent)
\be{prescription1E}
\tilde \mu_I^2 \propto M_P^2\,.
\ee
A second possibility is to choose the scale-invariant direction along
the dilaton field, i.e.
\begin{equation}
  \label{prescription2J}
  \mu_{II}^2\propto \xi_\chi \chi^2.
\end{equation}
When transformed to the Einstein frame it becomes  
\be{prescription2E}
\tilde \mu_{II}^2 \propto \frac{\xi_{\chi}\chi^2 M_P^2}{\xi_\chi \chi^2 + \xi_h h^2}\,,
\ee
and coincides with the prescription II of Ref.~\cite{Bezrukov:2009db} at the end of inflation.

In what follows we will use this ``minimal setup" for the analysis of the
radiative corrections. It will be  more convenient to work in the 
Einstein frame, where the coupling to gravity is minimal
and all non-linearities are moved to the matter sector.  The total action in the Einstein frame naturally divides into an Einstein- Hilbert (EH) part, a purely scalar piece involving only the Higgs and dilaton (HD) fields and a part corresponding to the chiral SM (CH) without the radial mode of the Higgs boson~\cite{Bezrukov:2009db,Dutta:2007st,Feruglio:1992wf}
  \begin{equation}
  S =S_\text{EH} +S_\text{HD}+ S_\text{CH}\,.
  \end{equation}
In the next section we estimate the contribution of the scalar sector to the effective inflationary potential, postponing the study of the chiral SM
to  Section \ref{cheffpot}. All the computations will be performed in flat spacetime, since the inclusion of gravity does not modify the
 results \footnote{We recall that, in the Einstein frame, the coupling among SM particles and gravity is minimal.}.

\subsection{Scalar contribution to the effective inflationary potential}
\label{scalareffpot}

Let us start by reminding that the initial value of the dilaton field has to be sufficiently
large to keep its present contribution to DE at the appropriate observational
level \cite{GarciaBellido:2011de}. The latter fact allows us to neglect
the exponentially suppressed contributions to the effective action stemming from 
$\tilde U_{\Lambda_0}$ in Eq.~\eqref{defin-pot}. As a result, the remaining corrections due 
to the dilaton field will emerge from its non-canonical kinetic term, whereas all the
 radiative corrections due to the Higgs field will emerge from the inflationary potential.

The construction of the effective action for the scalar sector of the theory is most
easily done in the following way: expand the action
\eqref{angul-theory} near the constant background of the dilaton and
the Higgs fields and drop the linear terms in perturbations. After
that, compute all the vacuum diagrams to account
 for the potential-type corrections and all the 
diagrams with external legs to account for the kinetic-type
corrections.

\subsubsection{Dilaton contribution}
Let us consider first the quantum corrections to the dilaton itself. 
Since our subtraction procedure respects the symmetries of the
classical action (in particular scale invariance, corresponding to the
shift symmetry of the dilaton field $r$ in the Einstein frame), no
potential terms for the dilaton can be generated. 
Thus, the loop expansion can only create two types 
of contributions, both stemming from its kinetic term. 
The first type are corrections to the
propagator of the field, and as we will show below they are
 effectively controlled by $(m_H/M_P)^{2k}$, with 
$m_H^2\equiv-\tilde U''(\phi)$ and $k$ the number 
of loops under consideration. The second type 
are operators with more derivatives of the field  
 suppressed by appropriate powers
of the scalar cut-off $M_P$. 
One should bear in mind that the appearance of these operators 
in the effective action is expected and
consistent. As discussed in the previous section, 
their presence does not affect the dynamics 
of the model, since the scalar cut-off is much larger than 
the characteristic momenta of the particles involved in
all physical processes throughout 
the whole history of the Universe. 

To demonstrate explicitly what
we described above, let us consider some of the associated diagrams.
Following the ideas of Ref.~\cite{Shaposhnikov:2008xi}, we perform the computations
in dimensional regularization in $D=4-2\epsilon$ dimensions. We avoid 
therefore the use of other regularizations schemes, such as cut-off 
regularization, where the scale invariance of the theory is badly 
broken at tree level\footnote{Similar arguments about the artifacts 
created by regularization methods that explicitly break scale 
invariance can be found for instance in Ref.~\cite{Bardeen:1995kv}.}. The magnitude 
of the corrections in dimensional regularization is of the order of the masses of the
 particles running in the loops, or in the case of the massless dilaton, its momentum. The
  structure of the corrections can be therefore guessed by simple power-counting and 
it becomes apparent already at the one-loop order. We get
\begin{figure}[H]
\centering
\includegraphics[scale=.75]{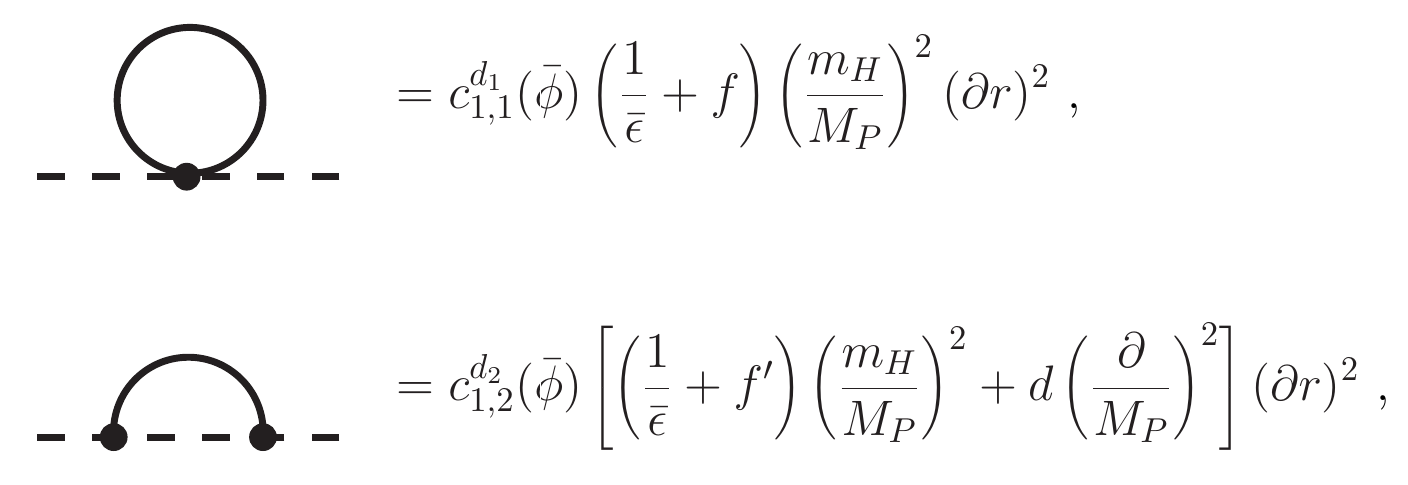}
\end{figure}
{\flushleft where} the Higgs and dilaton fields are represented by 
solid and dashed lines respectively. To keep the expressions as compact as possible
 we set $1/\bar \epsilon=1/\epsilon-\gamma+\log4\pi$ and denoted by $f \ \text{and} \ f'$ the finite parts 
of the diagrams, whose values depend on the normalization
point $\mu$. The higher-derivative operator in the second diagram is included for completion, but 
turns out to vanish accidentally  in this particular case.  Numerical factors are absorbed into
the background-dependent coefficients $c_{k,V}^{d_i}(\bar\phi)$,
which depend on the particular diagram $d_i$ under consideration, 
the number of loops $k$ and the number of vertices\footnote{We introduce the index  
$d_i$ to distinguish between the diagrams
with the same number of vertices but different combinations of hyperbolic 
functions that appear in higher loops.} $V$. Their values 
are always smaller than unity, and vary slightly with 
the background value $\bar\phi$. Their specific form of 
is presented in the Appendix B.

\begin{figure}[H]
\centering
\includegraphics[scale=.75]{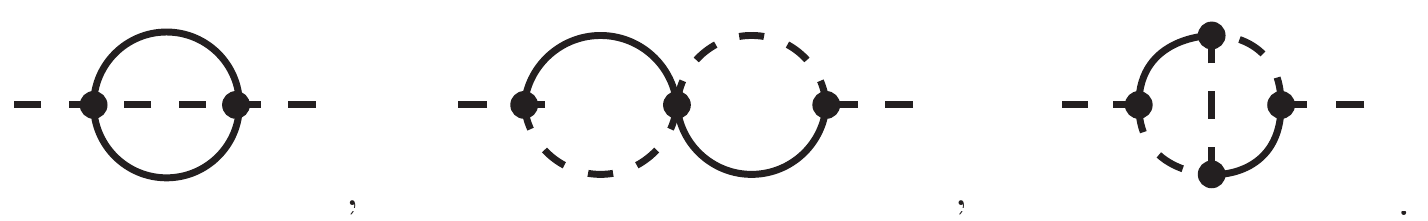}
\caption{Some of the two-loop diagrams
for the dilaton.}
\label{fig:dil-two-loop}
\end{figure}

In two-loops the situation is somehow similar.
The divergent (and finite) part of the corrections (consider
for example the diagrams presented in Fig.~\ref{fig:dil-two-loop}) 
is proportional to
\be{cor-2loop}
c_{2,V}^{d_i}(\bar\phi)
\left[\left(\dfrac{m_H}{M_P}\right)^4+
\left(\dfrac{m_H}{M_P}\right)^2
\left(\dfrac{\partial}{M_P}\right)^2+
\left(\dfrac{\partial}{M_P}\right)^4\right](\partial r)^2\ , \ \ \ V\le 4 \ . 
\ee
It is not difficult to convince oneself that this happens
in the higher order diagrams as well. The  
 structure of the corrections is therefore proportional to
\be{corr-dil1}
c_{k,V}^{d_i}(\bar\phi)
\left[\left(\dfrac{m_H}{M_P}\right)^{2k}+
\left(\dfrac{m_H}{M_P}\right)^{2k-2}
\left(\dfrac{\partial}{M_P}\right)^2+\ldots+
\left(\dfrac{m_H}{M_P}\right)^2\left(\dfrac{\partial}{M_P}\right)^{2k-2}+
\left(\dfrac{\partial}{M_P}\right)^{2k}\right](\partial r)^2\ ,
\ee
 up to  $\mathcal{O}(1)$ numerical factors.  Notice that some operators involving higher
 derivatives were already present at lower orders, but they reappear with extra suppression factors  $(m_H/M_P)^2$ on top of the scalar cut-off $M_P$. 
The corrections from diagrams with
gauge bosons and fermions running inside the loops are
given also by \eqref{corr-dil1}, by consistently replacing 
 $m_H$ by the mass of the particle considered. 

\subsubsection{Higgs contribution}
We now turn to the corrections to the Higgs field.
Once again we consider first the potential-type contributions.
The situation now is more complicated, since
the effective potential for the Higgs field $\phi$ will be modified
by terms stemming from the scale-invariant
part of the tree-level potential  \eqref{defin-pot} as well as from the non-canonical kinetic
term of the dilaton field $r$, with the latter starting from the second
order in perturbation theory. 

Let us start by considering the contributions due to the tree-level potential. To keep the 
notation as simple as possible, we express the scale-invariant
part of the potential \eqref{defin-pot} in the following compact form 
\be{compact-infl-pot}
\tilde U(\phi)= \lambda U_0\left(u_0+\sum_{n=1}^{2}u_{n}\cosh [2n
a\phi/M_P]\right), \hspace{.5cm}
U_0=\frac{M_P^4}{4\xi_h^2(1-\varsigma)^2} \ ,
\ee
where, for completion, we have explicitely recovered the $\alpha$ and $\beta$ dependence and defined
\be{coeffs}
u_0=c^2-c\sigma+\frac{3\sigma^2}{8}+\frac{3\beta'}{2} \ ,\
u_1=\frac{\sigma^2}{2}-c\sigma-2\beta' \ ,\
u_2=\frac{\sigma^2}{8}+\frac{\beta'}{2} \ , 
\ee
with 
\be{coefs-pot}
c=1+\frac{\alpha}{\lambda}\frac{1+6\xi_h}{1+6\xi_\chi} \ , 
\ \ \ \sigma=\varsigma+\frac{\alpha}{\lambda}
\frac{1+6\xi_h}{1+6\xi_\chi} \ , \ \ \ \  \beta'\equiv\frac{\beta}{\lambda}\left(\frac{1+6\xi_h}{1+6\xi_\chi}\right)^2 \ .
\ee
Expanding the field around its background value $\bar \phi$, we get
 \be{correct-pot}
\begin{aligned}
\tilde U(\bar \phi+\delta\phi)&= \lambda U_0
\sum_{n=1}^{2}u_n\sum_{l=0}^{\infty} 
\frac{\cosh^{(l)}[2n a\bar\phi/M_P]}{l!}
\left(\frac{2na\delta\phi}{M_P}\right)^l \\
&=\lambda U_0\sum_{n=1}^2\sum_{l=0}^\infty u_n\left[c_{n,l} 
\cosh[2n a\bar\phi/M_P]
\left(\frac{a\delta \phi}{M_P}\right)^{2l}+d_{n,l} 
\sinh[2na\bar\phi/M_P]
\left(\frac{a\delta \phi}{M_P}\right)^{2l+1}\right] \ ,
\end{aligned}
\ee
where $c_{n,l} \ \text{and} \ d_{n,l}$ account for numerical coefficients
and combinatorial factors. Since the theory is non-renormalizable, the
 perturbative expansion creates terms which do not have the 
same background dependence of the original potential. 
Up to numerical factors, the contributions turn out to be of the form\footnote{To maintain the expressions as  compact as possible we decided not to express the result in terms
of $m_H/M_P$.} 
\be{div-form}
\frac{\lambda^{i+j}M_P^4}{[4\xi_h^2(1-\varsigma)^2]^{i+j}}
 \left[g\left(\frac{1}{\epsilon}\right)+f_{i,j}\right]\sum_{n,m}u_n^iu_m^
j\cosh^i[2na\bar\phi/M_P]\sinh^j[2ma\bar\phi/M_P] \ ,
\ee
where
$f_{i,j}$ denotes the (finite) integration constant, and $g(1/\epsilon)$ 
is a function of the divergent terms. Note that if we set $\beta=0$,
 we make sure that terms which contribute to the cosmological 
constant \eqref{cosm-const} will not be generated 
by the loop expansion.  

By inspection of the structure of 
divergences, we can see that the leading corrections are 
those appearing with the lowest power in $\varsigma$. To gain 
insight on their contribution, we calculate the finite part of
 Eq.~\eqref{div-form}  for the maximal value of the hyperbolic functions.
 This corresponds to $\phi_\text{max}=\phi_0\equiv 
M_P/a\ \tanh^{-1}[\sqrt{1-\varsigma}]$. We get
\be{div-approx}
 \frac{\lambda^{i+j}}{[4\xi_h^2(1-\varsigma)^2]^{i+j}}f_{i,j}
\sum_{n,m}u_n^iu_m^j\cosh^i[2na\bar\phi/M_P]
\sinh^j[2ma\bar\phi/M_P]\Big\vert_{\bar\phi= 
\phi_\text{max}}\sim
\left(\frac{\lambda\varsigma}{4\xi_h^2}\right)^{i+j}f_{i,j} \ ,
\ee
which makes the corrections coming from the order $i+j+1$ negligible 
compared to the ones from $i+j$ order. In the last step we have simply
set $c=1,\ \sigma=\varsigma$, which, given the small value of 
the parameter $\alpha$ appearing in Eq.~\eqref{coefs-pot}, constitutes
a very good approximation.
\begin{figure}[t]
\centering
\subfigure[]{
\includegraphics[scale=.75]{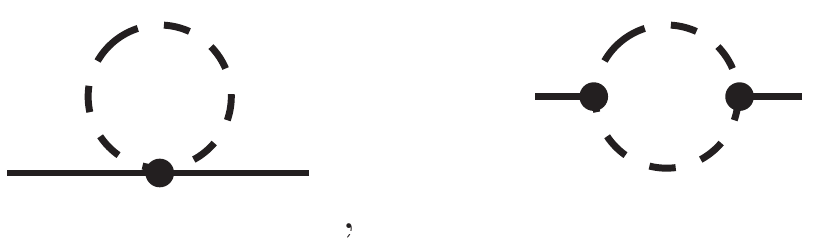}
\label{fig:oneloophig}}\\
\subfigure[]{
\includegraphics[scale=.75]{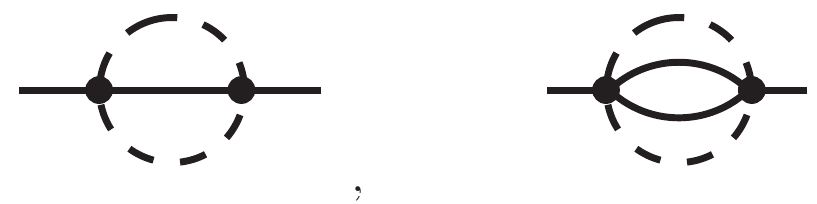}
\label{fig:twoloophig}}
\caption{Characteristic 
diagrams produced by the
non-canonical kinetic term of the dilaton field $r$.
Solid and dashed lines represent the Higgs and dilaton fields respectively.
The first one-loop diagram presented in (a)
vanishes in dimensional regularization due to the massless character 
of the dilaton field. On the other hand, the second diagram gives rise to 
higher derivative terms of the Higgs field. 
In (b) we consider two and three loop diagrams
which, apart from generating higher dimensional  operators,
 contribute to the effective potential once we amputate them.}
\end{figure}

As we mentioned earlier, potential-type corrections
to the Higgs field are also generated from diagrams
associated to the 
kinetic term of the dilaton $r$, starting from two loops.
This happens because the first order vacuum diagrams 
with dilaton running in the loop, vanish. If we consider higher loop 
diagrams, like those in Fig.~\ref{fig:twoloophig} but 
without momenta in the external legs, 
we see that even though the background
dependence of the corrections is complicated
due to the non-canonically normalized dilaton that 
runs inside the loops, their contributions
to the effective action are of the same order as 
 those in Eq.~\eqref{div-approx}. 

We now turn to the kinetic-type corrections to the Higgs
field. By that we mean corrections to the propagator, as well as terms with more derivatives of the field
suppressed by the scalar cut-off. 
The first type of contributions come only from the 
scale-invariant part of the potential given by \eqref{compact-infl-pot}
, when 
the momenta associated to the external legs are considered. 
It is not difficult to show that these are precisely of the same
form as those in \eqref{div-form}.
The second type of contributions, i.e. the higher dimensional
operators, are generated both from the Higgs potential at higher loops, as well as 
from the non-vanishing diagrams associated to the
 non-canonical kinetic term of the dilaton.
The terms we get are proportional to
\be{higherdim}
\frac{\partial^2}{M_P^2}(\partial\phi)^2 \ ,  \ \ \ 
\frac{\partial^4}{M_P^4}(\partial\phi)^2 \ \ldots \,,
\ee
and they can be safely neglected for the typical momenta involved in 
the different epochs of the evolution of the Universe.

Before moving on, we would like to comment on the 
appearance of mixing terms with derivatives of the fields.
These manifest themselves when we consider
diagrams with both fields in the external legs. 
They are higher dimensional operators, 
and it can be shown that they appear 
suppressed by the scalar cut-off of the theory, as before.

Since the kinetic-type operators do not modify
the dynamics, we will consider only potential-type corrections to estimate the change in the 
tree-level predictions of the model. At one-loop, the contribution of the scalar sector to the inflationary potential becomes \cite{Jackiw:1974cv}
\be{one-loop-correction}
\begin{aligned}
 \Delta \tilde U_{HD}&\simeq \frac{U_0}{64\pi^2}
\frac{\lambda a^4}{\xi_h^2(1-\varsigma)^2 }
\left(\frac{1}{\bar\epsilon}+f_{2,0}\right)\left[\varsigma^2
\frac{1+\cosh[4a\bar\phi/M_P]}{2}+\mathcal{O}(\varsigma^3)\right]\ ,
\end{aligned}
\ee
where we just kept the leading contribution in $\varsigma$. The finite part $f_{2,0}$ in the previous expression is given by 
\be{lfinite-part}
\begin{aligned}
f_{2,0}&=\frac{3}{2}-\log\left[\frac{-\tilde U''(\bar\phi)}{\mu^2}\right]
=\frac{3}{2}-\log\left[\frac{\lambda a^2 M_P^2}{\xi_h^2(1-\varsigma)^2\mu^2}\Bigg(
\varsigma \cosh[2a\bar\phi/M_P]+\mathcal{O}(\varsigma^2)\Bigg)\right] \ .
\end{aligned}
\ee
If we adopt the $\overline{MS}$ scheme, the remaining (logarithmic) corrections 
will be suppressed by an overall factor $\mathcal O(10^{-15})$ (apart from different powers of $\varsigma$) with respect to the tree-level 
potential \eqref{compact-infl-pot}. The quantum contribution of the scalar sector to the effective inflationary 
potential is therefore completely negligible and rather insensitive to the particular choice of the renormalization point $\mu$. This allows us to approximate the value of $\phi$ at the end of inflation by its classical value $\phi_f\simeq M_P/a \ \tanh^{-1} \left[\sqrt{1-\varsigma}\cos(2\times 3^{1/4}\sqrt{\xi_\chi})\right]$, and compute analytically the spectral tilt $n_s$ of 
primordial scalar perturbations, which turns out to be
\be{spec-tilt}
n_s(k_0)-1 \simeq -8\xi_\chi +\frac{\lambda\xi_\chi^2}{96\pi^2\xi_h^2}f_{2,0} \ ,
 \ \ \ \text{for} \ \ \   1\lesssim 4\xi_\chi N^*\ll 4N^* \ .
\ee
We see therefore that the correction to the tree-level result is controlled by the effective self-coupling of the Higgs field in the Einstein frame $\lambda/\xi_h^2$. The small value of this parameter makes the scalar radiative contribution completely negligible and thus hardly modify the consistency relation \eqref{functional-rel}. Note  however that there might be still a significant contribution to the inflationary potential coming from the SM particles, especially from those with a large coupling to the Higgs field. The study of this effect is the purpose of the next section. 

\subsection{Chiral SM contribution to the effective inflationary  potential.}\label{cheffpot}

The action for the SM fields during the inflationary stage is similar to that appearing in Higgs inflation 
\cite{Bezrukov:2009db} and takes the form of a chiral SM with a nearly decoupled Higgs field. Its 
contribution to the effective potential can be analyzed by the methods presented in Ref.~\cite{Bezrukov:2009db}.  
The one-loop contribution during inflation reads\footnote{We neglect the contribution \eqref{one-loop-correction}
 associated to the scalar sector, which, as shown in the previous section, turns out to be very small. }
\be{eq:Veff1}
  \Delta U_1 =
  \frac{6m_W^4}{64\pi^2}\left(\log\frac{m_W^2}{\mu^2}-\frac{5}{6}\right)
  + \frac{3m_Z^4}{64\pi^2}\left(\log\frac{m_Z^2}{\mu^2}-\frac{5}{6}\right)
  - \frac{3m_t^4 }{16\pi^2}\left(\log\frac{m_t^2}{\mu^2}-\frac{3}{2}\right)\,
\ee
where $m_W^2=g^2 h^2/2$, $m_Z^2=g^2h^2/2\cos^2\theta_W$ and
$m_t^2=y_t^2h^2/2$ stand for the effective $W, Z$ and top quark masses
in the Jordan frame.
The choice of the $\mu$ parameter here defines the renormalization prescription, as
described in the beginning of Sec.~\ref{sec:divergences}.  To retain
the possibility to use the RG equations to run between the
electroweak and inflationary scales we will write
$\mu^2=\frac{\hat{\mu}^2}{M_P^2}F(h,\chi)$.  Here the function $F(h,\chi)$
corresponds to the choice of the renormalization prescription and leads to
different physical results, while the parameter $\hat{\mu}$ plays the
role of the usual choice of momentum scale in the RG approach and
should disappear in the final result.
The conformal transformation to the Einstein frame $\Delta \tilde U_1 =\Delta U_1/\Omega^4$ acts only on the coefficients of the logarithmic terms in \eqref{eq:Veff1}, leaving their arguments completely unchanged.  We obtain therefore
 \be{eq:Veff2}
  \Delta \tilde U_1 =
  \frac{6\tilde m_W^4}{64\pi^2}\left(\log\frac{m_W^2}{\hat{\mu}^2F(h,\chi)/M_P^2}-\frac{5}{6}\right)
  + \frac{3\tilde m_Z^4}{64\pi^2}\left(\log\frac{m_Z^2}{\hat{\mu}^2F(h,\chi)/M_P^2}-\frac{5}{6}\right)
  - \frac{3\tilde m_t^4 }{16\pi^2}\left(\log\frac{m_t^2}{\hat{\mu}^2F(h,\chi)/M_P^2}-\frac{3}{2}\right)\,
\ee
where the Einstein-frame masses $\tilde m^2$ are proportional to the
effective vacuum expectation value of the Higgs field in the Einstein
frame\footnote{In particular we have $\tilde m_W^2(\phi)=\tilde
  m_Z^2(\phi) \cos^2\theta_w= g^2/2\cdot v^2(\phi)$ and $\tilde
  m_t^2(\phi)=y_t^2/2\cdot v^2(\phi)$.}, which is a slowly varying function during inflation,
\begin{equation}
  \label{eq:1}
  v^2(\phi) \equiv \frac{h^2}{\Omega^2}= \frac{M_P^2}{\xi_h(1-\varsigma)}
  \left(1-\varsigma\cosh^2\frac{a\phi}{M_p}\right).
\end{equation}
This fact
allows us to completely factor out the $\phi$ dependence in front of the logarithms in Eq.~\eqref{eq:Veff2} and perform the analysis  
below as if $v$ was a constant, $v\simeq M_P/\sqrt{\xi_h}$. 

Note that the explicit dependence on the  't Hooft-Veltman
normalization point $\hat{\mu}$ in Eq.~\eqref{eq:Veff1} is spurious
and is compensated by the running of the coupling constants
$\lambda(\hat{\mu})$, $\xi_h(\hat{\mu})$ in the
tree level part of the potential (see \cite{Bezrukov:2009db}).
 Once the RG running of the couplings is fixed,  it is convenient to
 choose the value of $\hat{\mu}$ in such a way that the logarithmic
 contribution   \ref{eq:Veff2}, for each given value $\phi$ of
 the Higgs field, is minimized, $\hat\mu^2\simeq \frac{y_t^2}{2}
 \frac{h^2}{F(h,\chi)/M_P^2}$.
 In that case, the RG enhanced (RGE) inflationary potential becomes
\begin{equation}
  \tilde U_\text{RGE}(\phi)=  \frac{\lambda(\hat{\mu}(\phi))}{4}
  \frac{M_P^4}{\xi^2_h(\hat{\mu}(\phi))(1-\varsigma)^2}
  \left(1-\varsigma\cosh^2\frac{a\phi}{M_p}\right)^2\,,
\end{equation}
which in fact suffices for practical purposes, with the corrections
form the 1-loop logarithms being rather small.

As discussed at the beginning of Section \ref{sec:divergences}, the
different choices of $\mu$ correspond to different subtraction  rules
and  produce different results.  In what follows we will consider the
two most natural  choices. The first one is associated to  the scale
invariant prescription \eqref{prescription1J}. The RG enhancement of 
the potential in this case dictates
\begin{equation}\label{PI}
  \hat\mu_\text{I}^2(\phi)
  = \frac{y_t^2}{2}\frac{M_p^2 h^2}{\xi_h h^2 + \xi_\chi\chi^2}
  = \frac{y_t^2}{2} v^2(\phi)\,,
\end{equation}
which is nothing else than the effective top mass in the Einstein frame. With this choice, the change in the shape of the 
potential is very small, given the insignificant variation of $v^2(\phi)$ during inflation. 
 The change in the inflationary observables $n_s$ and $r$ is therefore expected to be
 completely negligible. The second possibility that we will consider is associated to the prescription  \eqref{prescription2J}. In this 
 case the optimal choice of $\hat\mu$ is
\begin{equation}\label{PII}
  \hat\mu^2 _\text{II} (\phi)
  = \frac{y_t^2}{2}\frac{M_P^2 h^2}{\xi_\chi\chi^2}
  = \frac{y_t^2}{2} v^2(\phi)
    \frac{1-\varsigma}{\varsigma \sinh^2\left(a\phi/M_P\right)}\,,
\end{equation}
which, at the end of inflation, coincides with the effective top mass in the Jordan frame. This corresponds to the prescription II in Ref.~\cite{Bezrukov:2009db}. Note that contrary to the previous case, this choice strongly depends on the value of the $\phi$ field and noticeable contributions to the inflationary parameters are expected.  
\begin{figure}[!h]
  \begin{center}
\subfigure{\includegraphics[scale=0.6]{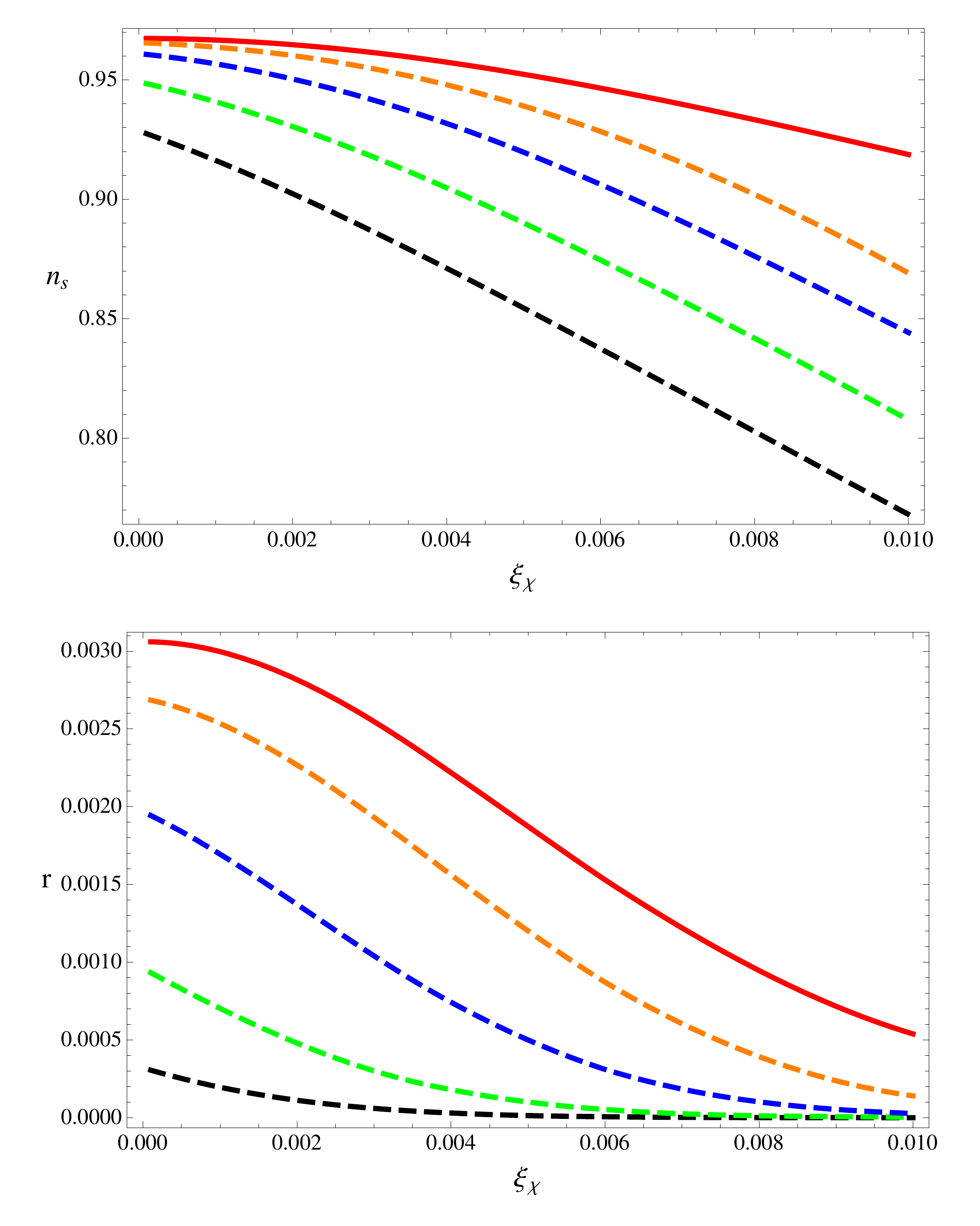}}
\fbox{\subfigure{\includegraphics[scale=0.8]{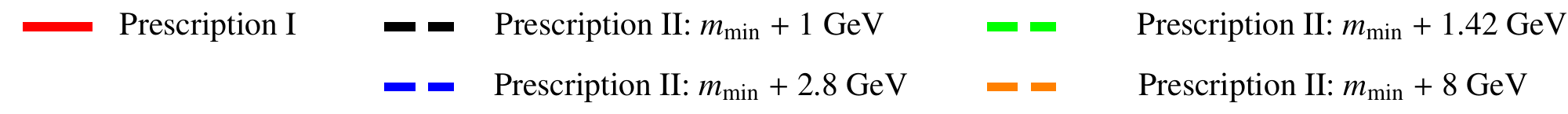}}}
  \end{center}
  \caption{The spectral index $n_s$ (top) and tensor to scalar ration $r$
    (bottom) as a function of the non-minimal coupling $\xi_\chi$.  The solid line corresponds to the
    quantization prescription I, which coincides with the tree level result. Dashed lines stand for the quantization choice II for different Higgs masses. The minimal Higgs boson
    mass $m_\text{min}$ can be obtained from Ref.~\cite{Bezrukov:2012sa}.}
  \label{fig:nsxichiRG}
\end{figure}

 The calculation proceeds now along the same lines as those in Ref.~\cite{Bezrukov:2009db}, using the tree level RG enhanced potential and the one loop correction. The addition of the two loop effective potential does not significantly modify the result. The numerical outcome for the two prescriptions is shown in Fig.~\ref{fig:nsxichiRG}. As expected, the inflationary observables computed with the first prescription coincide with the tree level result. The only effect of the quantum corrections is setting a minimal value for the Higgs mass. This turns out to be  $m_H>m_\text{min}$, with $m_\text{min}\simeq 129.5\pm\text{5 GeV}$ (for details on the latest calculations of this value see Ref.~\cite{Bezrukov:2012sa,Degrassi:2012ry}). After the end of inflation and preheating, the system is outside
the scale-invariant region and the fields settle
down to the minimum of the potential.  From the expansion of
the potential \eqref{darkener-pot}
 around the background, it is clear that all the
contributions to the effective action will be again suppressed by
powers of the exponent $e^{-\gamma r/M_P}$, in addition to powers
of $M_P$, not affecting therefore the predictions of the model
concerning the DE equation of state \eqref{reduc}. Taking into account the above results, we conclude that the quantum corrections
computed with the prescription I do not modify the classical consistency relation
\eqref{functional-rel} characterizing Higgs-Dilaton cosmology. On the other hand, the inflationary observables computed using the prescription II clearly differ from the tree level result, especially for Higgs masses close to the critical value $m_{min}$ at large $\xi_\chi$. Note that in this prescription, the recent observation of a light Higgs-like state \cite{:2012gu,:2012gk}, together with the present bounds on the spectral tilt $n_s$ \cite{Komatsu:2010fb}, further restrain the allowed $\xi_\chi$ interval. 

\section{conclusions}
\label{sec:conclusions}

The purpose of this paper was to study the self-consistency of the
Higgs-Dilaton cosmological model. We determined the field-dependent UV
cut-offs and studied their evolution in the different epochs
throughout the history of the Universe. We showed that the cut-off
value is higher than the relevant energy scales in the different
periods, making the model a viable effective field theory describing
inflation, reheating, and late-time acceleration of the Universe.
Since the theory is non-renormalizable, the loop expansion creates an
infinite number of divergences, something that may challenge the
classical predictions of the Higgs-Dilaton model. We argued that  this
is not the case if the UV-completion of the theory respects
scale-invariance  and the approximate shift symmetry for the dilaton
field. 

We computed within this framework the effective inflationary potential in the one-loop approximation and concluded that the dominant contribution comes from the chiral SM sector of the theory. We used two different regularizations prescriptions consistent with the symmetries of the model.  In the ``SI-prescription'' of  Ref.~\cite{Shaposhnikov:2008xi}, with a field-dependent normalization point proportional to the effective Planck scale in the Jordan frame, the effective potential turns out to coincide with the tree level one. This leaves practically intact the consistency relation \eqref{functional-rel} which connects the inflationary spectral tilt to the deviation of the DE equation of state from a cosmological constant. This relation is however modified if the normalization point is chosen only along the dilaton's direction, especially for Higgs masses near the critical value $m_\text{min}\simeq 129.5\pm\text{5 GeV}$, which is amazingly close to the mass of the recently observed Higgs-like particle at the LHC \cite{:2012gu,:2012gk}. In the lack of a Planck scale UV completion, the proper choice of the normalization point $\mu$ can only be elucidated by improving the precision of the cosmological and particle physics observables.

\begin{acknowledgements}

GKK would like to thank Kostas Farakos for numerous discussions. 
JR thanks Juan Garc\'ia-Bellido for valuable comments.
This work was supported in part by the Swiss National Science
Foundation, the Tomalla Foundation and the Greek State Scholarship Foundation through the
LLP-ERASMUS program.  	
\end{acknowledgements}

\begin{appendix}\label{appendix1}
\section{Einstein frame cut-offs}

In this appendix we briefly discuss the computation of the effective cut-off  in the Einstein frame. As before, the cut-off is understood as the energy at which perturbative unitarity is violated and not necessarily as the onset of new physics. As shown in Eq. \eqref{einst-theory}, the gravitational part of the action in the transformed frame takes the usual Einstein-Hilbert form, which allows us to directly identify the gravitational cut-off with the reduced Planck mass $M_P$.  The cut-off associated to the gauge sector can be also easily determined by looking at the scattering of gauge bosons with longitudinal polarization. Since the kinetic terms for the gauge fields are invariant under the conformal rescaling, the only modification comes through their coupling to the Higgs field $h$. The interaction under consideration can be schematically written as
\be{coup-app}
g^2 h^2 W_\mu^+ W^{-\mu}\rightarrow g^2 \frac{h^2}{\Omega^2}\tilde W_\mu^+ \tilde W^{-\mu} \ .
\ee
where we have rescaled the gauge boson fields in the Einstein frame with the corresponding conformal weight, $\tilde W^{\pm}=W^{\pm}/\Omega$.
Expanding \eqref{coup-app} around the background value of the Higgs field, $h\rightarrow \bar h +\delta h$, we find the following interaction term
\be{}
g\, \frac{m_W}{\bar\Omega^2} \tilde W_\mu^+ \tilde W^{-\mu}\delta h \ , 
\ee
where $m_W\sim g \bar h$ is the mass of the $W$ bosons in the Jordan frame and the conformal factor $\bar\Omega$ depends now on the background values of the Higgs and dilaton fields.
Taking into account the canonically normalized perturbations of the Higgs field  \eqref{redefinition-fields-einstein}, together with the unitarity of the S-matrix, we find that the cut-off scale associated to the gauge sector is given by 
\be{einst-gauge-app}
\tilde \Lambda_G\simeq \bar\Omega^{-1}\sqrt{
\frac{\xi_\chi\bar\chi^2(1+6\xi_\chi)+\xi_h\bar
h^2(1+6\xi_h)}{6\xi_h^2}} \ .
\ee
For the two limiting cases discussed in Section \ref{sec:gauge-cut-off} the previous expression becomes
\be{lim-gauge} 
\tilde \Lambda_G\simeq \Bigg\{ 
\begin{array}{cl} \frac{M_P}{\sqrt{\xi_h}}
&\mbox{for} \ \xi_\chi\bar\chi^2\ll \xi_h \bar h^2 \ ,\vspace{.1cm}
 \\
\frac{M_P}{\xi_h} &\mbox{for} \
\xi_\chi\bar\chi^2\gg \xi_h \bar h^2 \ .\\
\end{array} 
\ee 
where we have identified $\sqrt{\xi_\chi}\chi=M_P$. As expected, the gauge cut-off in the Einstein frame  is nothing else that the conformal rescaling of the Jordan frame cut-off, $\tilde\Lambda_G=\Lambda_G/\Omega$.

The computation of the scalar cut-off in the Einstein frame is more complicated than in the single field case \cite{Bezrukov:2010jz}. Although all the non-linearities of the initial frame are moved to the matter sector of the theory,  the existence of non-minimal couplings to gravity give rise to a non-trivial kinetic mixing for the scalar fields in the Einstein frame (cf. Eq. \eqref{kinetic}). This fact substantially complicates the treatment of the problem in terms of the original $(h,\chi)$ variables, especially in the high energy region. Therefore, in order to compute the scalar cut-off at large energies, we choose to recast the kinetic terms \eqref{kinetic} in a diagonal form by means of the angular variables defined in \eqref{var-r-theta}. Expanding the resulting inflationary potential\footnote{Equivalently we could consider higher order terms arising from the non-canonical kinetic term of the dilaton.} in  Eq. \eqref{defin-pot} around the background value of the Higgs field $\bar\phi$ we obtain a  series of  terms of the form (cf. Eq \eqref {correct-pot})
\be{exp-pot-app}
c_{n,l} 
\cosh[2n a\bar\phi/M_P]
\left(\frac{a\delta \phi}{M_P}\right)^{2l}+d_{n,l} 
\sinh[2na\bar\phi/M_P]
\left(\frac{a\delta \phi}{M_P}\right)^{2l+1}\ . 
\ee
The scalar cut-off  during inflation and reheating can be directly read from the previous expression. Note however that a direct comparison of the previous result with those obtained in the Jordan frame is only possible in some limiting cases. The angular perturbation $\delta\phi$ depends on both of the original field perturbations and only coincides with the Higgs perturbation $\delta h$ in the very high energy regime. Indeed, at the beginning of inflation\footnote{The background value of the field $\phi$ is very close to zero. Remember that $\phi$ is defined as $\phi=\phi_0-\vert\phi'\vert$.} the angular dependence  on the background field in Eq. \eqref{exp-pot-app} becomes negligible. We are left therefore with a series of higher order operators suppressed by the reduced Planck mass  $M_P$, which coincides with the conformally transformed Jordan frame cut-off in the corresponding regime, $\tilde \Lambda\simeq \Lambda/\Omega \simeq \sqrt{\xi_h}h/\Omega$.

The determination of the scalar cut-off in the low-energy regime, $\xi_h h^2\ll  \xi_\chi\chi^2$, is also non-trivial, since the field redefinition  \eqref{var-r-theta} is no longer applicable. Fortunately, the kinetic mixing between the Higgs and dilaton fields can be neglected at low energies and Eq. \eqref{kinetic} simplifies to
\be{kinetic-app-le}
\tilde K(\chi,h) \simeq (\partial \chi)^2+\left(1+\frac{\xi_h^2h^2}{M_P^2}\right)(\partial h)^2 \ ,\ee
where again we identified  $\sqrt{\xi_\chi}\chi=M_P$.
The kinetic term for the Higgs field can be recast into canonical form in terms of 
\be{chvar-app-le}
\hat h= h\left(1+\frac{\xi_h^2 h^2}{M_P^2}+\ldots\right)=h\left(1+\sum_{n=1}c_n\left(\frac{\xi_h^2h^2}{M_P^2}\right)^n\right) \ ,
\ee
where $c_n$ are numerical factors. Inverting the above relation and plugging it to the potential in this limit
\be{pot-le-app}
\tilde U(h)\simeq \frac{\lambda}{4}h^4 \ ,
\ee
we see that the cut-off is proportional to $M_P/\xi_h$, in agreement with the Jordan frame result\footnote{Notice that in the low energy regime the conformal factor is approximately equal to one.}.
\clearpage

\section{Feynman rules for the dilaton}\label{appendix2}
In this appendix we gather the Feynman rules as well as the expressions for the coefficients appearing in the one-loop diagrams in Section \ref{scalareffpot}. We denote with a dashed (solid) line the dilaton (Higgs) and perform the calculations in dimensional regularization in $D=4-2\epsilon$ dimensions.  After expanding the fields around their background values and normalizing the kinetic term for the dilaton, we find the following Feynman rules stemming from its kinetic term   
\begin{figure}[!h]
\centering
\includegraphics[scale=.75]{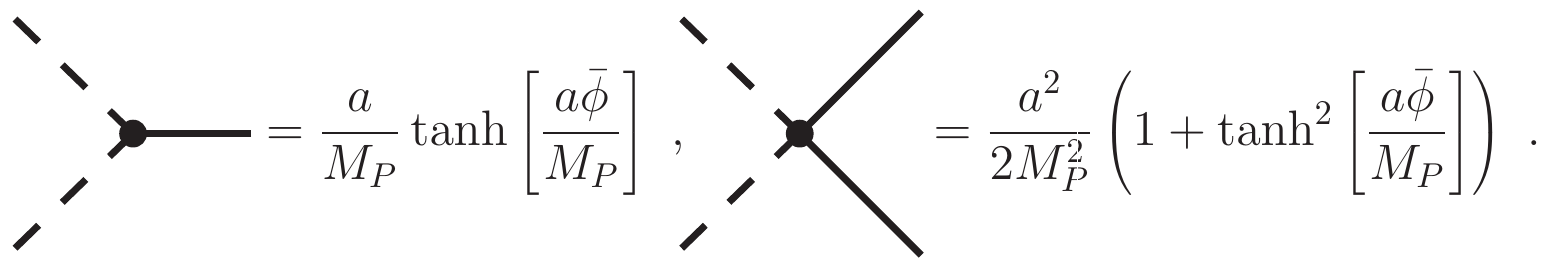}  
\end{figure}

Using the above expression, we can calculate the coefficients appearing in the different diagrams. Let us start by considering the simplest diagram $d_1$ . We obtain

\begin{figure}[!h]
\centering 
\includegraphics[scale=.75]{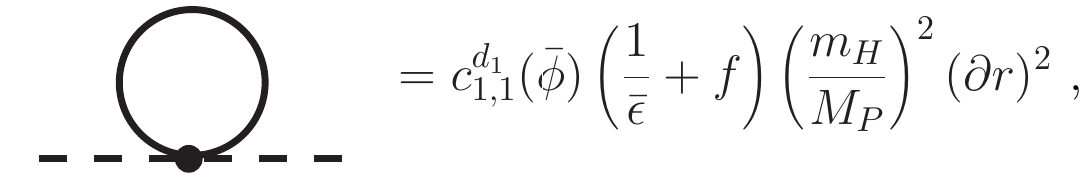}
\end{figure}

{\flushleft with} $1/\bar\epsilon=1/\epsilon-\gamma+\log4\pi$, and  
\be{appeq1} 
c^{d1}_{1,1}(\bar\phi)=\frac{a^2}{64\pi^2}\left(1+\tanh^2\left [\frac{a\bar\phi}{M_P}\right] \right) \ ,  \ \ \
f=-\log\left[\frac{m_H^2}{\mu^2}\right] \ .
\ee
Let us move to the more complicated diagram $d_2$. We find

\begin{figure}[!h]
\centering
\includegraphics[scale=.75]{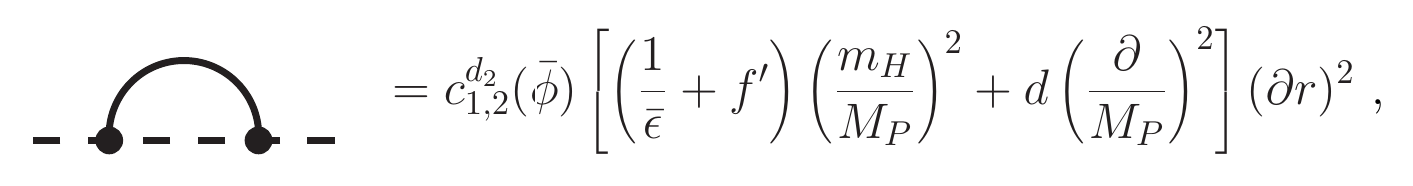}
\end{figure}

{\flushleft where}
\be{appeq2} 
c^{d2}_{1,2}(\bar\phi)=\frac{a^2}{16\pi^2}\tanh^2\left [\frac{a\bar\phi}{M_P}\right] \ ,  \ \ \ 
f'=\frac{1}{2}-\log\left[\frac{m_H^2}{\mu^2}\right]  \ \ \ \text{and} \ \ \ d=0 \ .
\ee
Note that in this particular diagram, the coefficient $d$ is coincidentally zero. As we argued in Section \ref{scalareffpot},  this kind 
of terms are expected to appear by simple power-counting arguments in higher-loop diagrams.  We see that in both diagrams, 
for the maximal value of the hyperbolic tangent,  the corrections are suppressed by loop factors as well as powers of $M_P$.
\end{appendix}

\end{document}